\documentclass[useAMS,usenatbib]{mn2e}
\usepackage{longtable}
\usepackage{graphicx}

\newcommand{\Porb}{\mbox{$P_{\mathrm{orb}}$}}
\newcommand{\Psh}{\mbox{$P_{\mathrm{sh}}$}}
\newcommand{\Teff}{\mbox{$T_{\mathrm{eff}}$}}
\newcommand{\Msun}{\mbox{$M_\odot$}}

\newcommand{\Ion}[2]{#1{\,\scriptsize #2}}

\newcounter{tref}
\newcommand{\tbr}{$^{\arabic{tref}}$\stepcounter{tref}}
\newcommand{\tbc}{\arabic{tref}\stepcounter{tref}}

\title[SDSS CVs: a period minimum spike unveiled]{SDSS unveils a
 population of intrinsically faint cataclysmic variables at the
  minimum orbital period}

\author[B.T. G\"ansicke et al.]{
B.T. G\"ansicke$^1$,
M. Dillon$^{1}$,
J. Southworth$^1$,
J.R. Thorstensen$^2$,
P. Rodr{\'i}guez-Gil$^{3,4}$, \newauthor
A. Aungwerojwit$^{1,5}$,
T.R. Marsh$^1$,
P. Szkody$^6$,
S.C.C. Barros$^1$,
J. Casares$^4$,\newauthor
D. de Martino$^7$,
P.J. Groot$^8$, 
P. Hakala$^9$,
U. Kolb$^{10}$,
S.P. Littlefair$^{11}$,\newauthor
I.G. Mart{\'\i}nez-Pais$^{4,12}$,
G. Nelemans$^8$,
M.R. Schreiber$^{13}$
\medskip\\
$^1$ Department of Physics, University of Warwick, Coventry CV4 7AL,
UK \\
$^2$ Department of Physics and Astronomy, Dartmouth College, 6127 Wilder
Laboratory, Hanover, NH 03755-3528, USA\\
$^3$ Isaac Newton Group of Telescopes, Apartado de correos 321, E-38700 Santa Cruz de la Palma, Tenerife, Spain\\
$^4$ Instituto de Astrof{\'\i}sica de Canarias, 38200 La Laguna,
Tenerife, Spain\\
$^5$ Department of Physics, Faculty of Science, Naresuan University, 
Phitsanulok, 65000, Thailand\\
$^6$ Astronomy Department, University of Washington, Seattle, WA98195, USA\\
$^7$ INAF - Osservatorio di Capodimonte, Via Moiariello 16, 80131 Napoli, Italy\\
$^8$ Department of Astrophysics, IMAPP, Radboud University Nijmegen,
  P.O. Box 9010, 6500 GL Nijmegen, The Netherlands\\
$^9$ Tuorla Observatory, University of Turku, Vaisalantie 20,
  FIN-21500 Piikkio, Finland\\
$^{10}$ Department of Physics and Astronomy, The Open University,
  Walton Hall, Milton Keynes MK7 6AA, UK\\
$^{11}$ Department of Physics and Astronomy, University of Sheffield, S3 7RH\\
$^{12}$ Departamento de Astrof{\'\i}sica, Universidad de La Laguna, E-38206 La
Laguna, Tenerife, Spain\\
$^{13}$ Departamento de F\'isica y Astronom\'ia, Facultad de Ciencias, Universidad 
de Valpara\'iso, Valpara\'iso, Chile 
}

\begin{document}

\date{Accepted 2005 Month day. Received 2005 Month day; in original form 2005 Month day}

\pagerange{\pageref{firstpage}--\pageref{lastpage}} \pubyear{2005}

\maketitle

\label{firstpage}

\begin{abstract}
We discuss the properties of 137 cataclysmic variables (CVs) which are
included in the Sloan Digital Sky Survey (SDSS) spectroscopic data
base, and for which accurate orbital periods have been measured. 92 of
these systems are new discoveries from SDSS and were followed-up in
more detail over the past few years. 45 systems were previously
identified as CVs because of the detection of optical outbursts and/or
X-ray emission, and subsequently re-identified from the SDSS
spectroscopy. The period distribution of the SDSS CVs differs
dramatically from that of all the previously known CVs, in particular
it contains a significant accumulation of systems in the orbital
period range 80--86\,min. We identify this feature as the elusive
"period minimum spike" predicted by CV population models, which
resolves a long-standing discrepancy between compact binary evolution
theory and observations.  We show that this spike is almost entirely
due to the large number of CVs with very low accretion activity
identified by SDSS. The optical spectra of these systems are dominated
by emission from the white dwarf photosphere, and display little or no
spectroscopic signature from the donor stars, suggesting very low-mass
companion stars. We determine the average absolute magnitude of these
low-luminosity CVs at the period minimum to be
$<M_g>=11.6\pm0.7$. Comparison of the SDSS CV sample to the CVs found
in the Hamburg Quasar Survey and the Palomar Green Survey suggests
that the depth of SDSS is the key ingredient resulting in the
discovery of a large number of intrinsically faint short-period
systems.
\end{abstract}


\section{Introduction}
Compact binary stars (CBs) are central to a variety of astrophysical
contexts such as short $\gamma$-ray bursts, which are thought to arise
from neutron star mergers \citep{paczynscki86-1, gehrelsetal05-1,
  bergeretal05-1}; supernovae type Ia, whose progenitors are white
dwarf binaries \citep{woosleyetal86-1, whelan+iben73-1,
  ruiz-lapuenteetal04-1}; or stellar black holes
\citep{casaresetal92-1}. Despite the fundamental importance of CBs on
a wide range of astronomical scales, our understanding of their
evolution is rather fragmentary. Major uncertainties lie in the common
envelope phase, when the more massive of the two stars evolves into a
red giant and engulfs the other one \citep[e.g.][]{nelemans+tout05-1},
and in the subsequent orbital angular momentum losses such as magnetic
stellar wind braking \citep{verbunt+zwaan81-1, kalogeraetal98-1,
  justhametal06-1}.

Cataclysmic variables (CVs) are compact binaries containing a white
dwarf accreting from a Roche-lobe filling main-sequence, slightly
evolved, or brown-dwarf donor and represent an abundant class of CBs
that are well-suited for observational tests of CB evolution
theory. However, while detailed CV population models have been
calculated \citep[e.g.][]{kolb93-1, politano96-1, howelletal01-1}, and
more than 600 CVs have been observed in some detail
\citep{ritter+kolb03-1}, the disagreement between theory and
observations has been a continuous source of frustration for the past
two decades.

CVs with main-sequence or slightly evolved donors evolve towards
shorter orbital periods through the loss of angular momentum by 
magnetic braking \citep{verbunt+zwaan81-1, rappaportetal83-1} and
gravitational radiation \citep{faulkner71-1, paczynski+sienkiewicz81-1}
until the mass of the donor becomes too low to sustain
hydrogen-burning, and the donor starts to become degenerate.  At this
point, the orbital period evolution reverses sign so that the period
increases with time, and hence a very strong prediction of CV
evolution theory is the existence of a minimum orbital period
\citep{rappaportetal82-1, paczynski+sienkiewicz83-1}. Early estimates
predicted the period minimum to occur around
$P_\mathrm{min}\simeq60-80$\,min \citep{rappaportetal82-1,
  paczynski+sienkiewicz81-1}, whilst more recent work using improved physics
in the modelling of low-mass stars results in
$P_\mathrm{min}\simeq65-70$\,min \citep{kolb+baraffe99-1,
  howelletal01-1}. While a sharp minimum is detected in the period
distribution of the currently known CVs \citep[e.g.][]{knigge06-1}, it
is found at $\simeq76$\,min. This value is significantly longer than the
predicted value, which could imply additional sinks of angular
momentum loss besides gravitational wave radiation
\citep[e.g.][]{patterson98-1}, and/or problems in the understanding
the structure of mass-losing low-mass stars. 

Since the detection probability of a system at a given range in
orbital period (\Porb) is proportional to the time it requires to
evolve through that range, $N(\Porb)\propto 1/|\dot {\Porb}|$, a
significant accumulation of CVs at the minimum period is
expected. This accumulation is often dubbed \textit{the period minimum
  spike} \citep{kolb+baraffe99-1}. Confronting theory and
observations, it is clear that such a spike is absent in the period
distribution of the known CVs \citep{patterson98-1, gaensickeetal02-2,
  knigge06-1}.  This absence of the predicted pile-up of systems near
the minimum period among the observed CVs has been one of the most
intensively debated discrepancies between CV evolution theory and
observations \citep{patterson98-1, kolb+baraffe99-1, kingetal02-1,
  barker+kolb03-1, willemsetal05-1}, casting doubts on our
understanding of compact binary evolution in general.

A major uncertainty on the observational side has been the fact that
the known sample of CVs is a rather mixed bag of systems discovered by
a variety of methods, such as variability, X-ray emission, or blue
colour. It is therefore subject to complex selection effects
\citep{gaensicke05-1}, and should not be expected to provide a genuine
representation of the intrinsic properties of the galactic CV
population. In particular, all conventional CV discovery methods
favour systems with intermediate mass transfer
rates\footnote{Ironically, non-magnetic CVs with high mass transfer
rates are also missed by the main identification methods, i.e. the
detection of large-amplitude variability or X-ray emission. The reason
for this is that these systems contain accretion discs in a hot,
stable state with no occurrence of outbursts, and little X-ray emission
from the boundary layer on the white dwarf. A substantial number of
bright high mass transfer systems has only been found in blue-colour
and/or emission line surveys, such as the Palomar Green Survey
\citep[e.g.][]{thorstensenetal91-2} or the Hamburg Quasar Survey
\citep{aungwerojwitetal05-1, rodriguez-giletal07-2}. A recent example
for previously mis-identified bright CVs is the 11th magnitude system
LS\,IV-08$^{\circ}$3, which was thought to be an OB star, but is in fact
a novalike variable \citep{starketal08-1}.}, i.e. systems
that are characterised by frequent outbursts or large X-ray
luminosities. Consequently, the observed sample of CVs is likely to
under-represent systems with low mass transfer rates, as they rarely
undergo outbursts and are X-ray faint. Those are, however, exactly the
properties predicted by population models for old, short-period
systems, which are thought to make up the majority of all CVs
\citep{kolb93-1, howelletal97-1}.

The CV sample identified in the Sloan Digital Sky Survey
\citep{szkodyetal02-2, szkodyetal03-2, szkodyetal04-1, szkodyetal05-1,
  szkodyetal06-1, szkodyetal07-2, szkodyetal09-1} has the potential to
overcome many of the limitations of existing samples, as it extends to
fainter objects and covers a wider range in colour space than any
previous optical survey. In a series of earlier papers we have
explored the properties of individual SDSS CVs
\citep[e.g.][]{szkodyetal03-3, wolfeetal03-1, peters+thorstensen05-1,
  gaensickeetal06-1, littlefairetal06-1, southworthetal06-1,
  southworthetal07-2, dillonetal08-1}.  Here, we compile accurate
orbital period measurements for 137 SDSS CVs, and show that the period
distribution of these systems differs dramatically from all previously
examined samples. In particular, a clear accumulation of systems in
the period range 80--86\,min is observed, consistent with the
prediction of a period minimum spike made by the population
models. Furthermore, we show that the bulk of the systems in this
spike display distinctly different properties from the average
short-period ($\la120$\,min) CV.

\section{Accurate orbital period measurements of SDSS cataclysmic variables}
For the purpose of quantitative comparison between the orbital period
distribution of an observed sample of CVs and that of a population
model, care has to be taken over the period measurements included in
the observed distribution. For the convenience of the reader
unfamiliar with the observational techniques, we briefly outline the
caveats of CV orbital period determinations before describing the
details of our compilation of accurate SDSS CV period measurements.

Orbital period determinations of CVs are generally achieved by one of
two methods: time-resolved spectroscopy probing for radial velocity
variations of at least one component in the CV, and time-resolved
photometry probing for coherent variability of the system's
brightness. Both methods can yield straightforward and unambiguous
results, e.g. using the radial velocities measured from sharp
absorption lines associated with the donor star, or from the detection
of eclipses in the light curve. However, in a large fraction of CVs,
more care has to be taken in the interpretation of the observations,
as e.g. radial velocities measured from emission lines in
low-resolution spectroscopy may reflect the unresolved motion of
different components within the CV, or red-noise and flickering in
light curves may mimic a periodic brightness modulation. Such
systematic problems can often, but not always, be overcome by
increasing the spectral and/or temporal resolution, signal-to-noise
ratio, and the total quantity of data accumulated for a given
system. Given the typical oversubscription on telescopes, it is an
unfortunate fact that observers have to strike a difficult balance
between the number of systems that can be followed up, and the quality
of parameters determined for each of them. 

\begin{table*}
\caption{CVs identified by SDSS (\textit{new SDSS CVs}) with reliable
  orbital period determinations. Orbital periods followed by '*' are
  calculated from Eq.\,\ref{e-psh-porb}, but assuming a somewhat more
  pessimistic error of 2\,min. The $g$ magnitude refers to the
  typical state of the system, i.e. quiescence for dwarf novae and
  high state for novalike variables. The CV subtype is given as
  DN\,=\,dwarf nova, i.e. at least one optical outburst has been
  detected; AM\,=\,AM\,Her star, i.e. the system shows characteristics
  of a strongly magnetic CV, IP\,=\,intermediate polar, i.e. coherent
  short-period variability related to the white dwarf spin is
  detected; NL\,=\,novalike variable, i.e. the system is in a
  persistent state of high mass transfer; CV\,=\,no confirmed CV
  subtype; EC\,=\,eclipsing. Two additional flags are given,
  WD\,=\,the SDSS spectrum is dominated by the white dwarf,
  RASS\,=\,the system has been detected in the ROSAT All Sky Survey.}
\label{t-porbs}
\setcounter{tref}{1}
\begin{tabular}{lllllll}
\hline
SDSSJ   &  $g$ & $P_\mathrm{orb}$\,[min] & Type & WD & RASS & Reference\\
\hline
003941.06+005427.5 & 20.6 & 91.50(16)           & CV    & Y & N & IV,\tbc \\
004335.14-003729.8 & 19.8 & 83.39(8)            & CV    & Y & N & III,\tbc \\
005050.88+000912.7 & 20.4 & 80.3(2.2)           & CV    & Y & N & IV,\tbc  \\
013132.39-090122.3 & 18.3 & 81.54(13)           & CV    & Y & N & II,3  \\
013701.06-091234.9 & 18.7 & 79.71(1)            & DN    & Y & N & II,\tbc  \\
015151.87+140047.2 & 20.3 & 118.68(4)           & DN    & N & N & I,\tbc   \\
015543.40+002807.2 & 15.4 & 87.143517(1)        & AM/EC & N & Y & I,\tbc,\tbc,\tbc \\
023322.6+005059.50 & 19.9 & 96.08(9)            & IP:   & N & N & I,\tbc    \\
031051.66-075500.2 & 22.3 & 95.9(2.0)*          & DN    & N & N & II,\tbc \\
032855.00+052254.2 & 18.0 & 121.97(25)          & AM    & N & N & VI    \\
040714.78-064425.1 & 17.8 & 245.045(43)         & DN/EC & N & N & II,\tbc  \\
074531.92+453829.6 & 19.0 & 76.00(16)           & CV    & Y & N & V,1\\
074640.62+173412.8 & 21.1 & 93.5(2.0)*          & DN    & N & N & V,10\\
075059.97+141150.1 & 19.1 & 134.1530(39)        & CV/EC & N & N & VI,1\\
075240.45+362823.2 & 17.7 & 164.4(3.0)          & AM    & N & N & II,\tbc \\
075443.01+500729.2 & 17.3 & 205.965(14)         & NL/EC & N & N & V,3   \\
075507.70+143547.6 & 18.3 & 84.760(58)          & CV    & Y & N & VI,1  \\
075939.79+191417.3 & 18.2 & 188.45(80)          & CV    & N & N & V,1   \\
080215.38+401047.1 & 16.7 & 221.62(4)           & NL    & N & N & II,\tbc \\
080434.20+510349.2 & 17.9 & 84.96(4)            & DN    & Y & N & V,\tbc,\tbc,\tbc \\
080534.49+072029.1 & 18.5 & 330.8(9)            & CV    & N & N & VI,\tbc \\
080846.19+313106.0 & 19.4 & 296.45(75)          & DN    & N & N & III,13 \\
080908.39+381406.2 & 15.6 & 193.015(12)         & NL/EC & N & N & II,\tbc \\
081256.85+191157.8 & 15.8 & 230.4(3)            & NL    & N & N & V,17 \\
081207.63+131824.4 & 19.3 & 116.8(2.0)*         & SU    & N & N & VI,10\\
081321.91+452809.4 & 18.3 & 416.2(6)            & DN    & N & N & I,\tbc \\
081352.02+281317.3 & 17.1 & 175.11(56)          & CV    & N & N & IV,17\\
082409.72+493124.4 & 19.3 & 95(3)               & DN    & N & N & I,5,\tbc \\
083845.23+491055.4 & 19.6 & 99.7(2.0)*          & DN    & N & N & I,10\\
084400.10+023919.3 & 18.3 & 298.1(7)            & CV    & N & N & II,17 \\
090016.56+430118.2 & 18.9 & 301.492(78)         & CV    & N & N & III,1 \\
090103.93+480911.1 & 19.3 & 112.147927(50)      & DN/EC & N & N & II,5 \\
090350.73+330036.1 & 18.8 & 85.065902(13)       & CV/EC & N & N & IV,13,\tbc \\
090403.48+035501.2 & 19.2 & 86.000(8)           & CV    & Y & N & III,\tbc\\
091127.36+084140.7 & 19.7 & 295.74(22)          & CV    & N & N & IV,9\\
091945.11+085710.0 & 18.2 & 81.3(2)             & CV    & Y & N & IV,5,17 \\
092009.54+004244.9 & 17.5 & 212.944(11)         & CV/EC & N & N & II,\tbc \\
092122.83+203857.0 & 19.8 & 84.2409(36)         & AM    & N & N & VII,1,\tbc\\
092444.48+080150.9 & 19.3 & 131.2560(67)        & AM:/EC& N & N & IV,1\\
093249.56+472523.0 & 17.8 & 95.47669(11)        & CV/EC & N & N & III,13,\tbc\\
100515.38+191107.9 & 18.2 & 107.6(2.0)*         & DN    & Y & N & VII,10\\
100658.40+233724.4 & 18.3 & 267.7163(30)        & CV/EC & N & N & VI,1 \\
103100.55+202832.2 & 18.3 & 83.2(2.3)           & AM    & N & N & \tbc\\
103533.02+055158.3 & 18.8 & 82.0897(3)          & CV/EC & Y & N & V,9,21,\tbc\\
110014.72+131552.1 & 18.7 & 94.5(2.0)*          & DN    & N & Y & V,10\\
115207.00+404947.8 & 19.3 & 97.4(4)             & CV/EC & Y & N & VI,1\\
121209.31+013627.7 & 18.0 & 88.428(1)           & AM    & Y & N & \tbc,\tbc,\tbc\\
121607.03+052013.9 & 20.1 & 98.82(16)           & CV    & Y & N & III,9 \\
122740.82+513924.9 & 19.1 & 90.64859(1)         & DN/EC & Y & N & V,21,\tbc\\
123813.73-033932.9 & 17.8 & 80.52(50)           & CV    & Y & N & II,\tbc\\
124426.25+613514.5 & 18.8 & 142.9(2)            & CV    & N & Y & III,5  \\
125023.84+665525.4 & 18.7 & 84.5793893(63)      & CV/EC & N & N & II,5 \\
132411.57+032050.4 & 22.1 & 158.72(9)           & AM    & N & N & III,1,\tbc \\
132723.38+652854.3 & 17.8 & 196.772(89)         & NL/EC & N & N & II,\tbc \\
133941.11+484727.5 & 17.6 & 82.524(24)          & CV    & Y & N & IV,\tbc \\
143317.78+101123.3 & 18.6 & 78.106578(3)        & CV/EC & Y & N & IV,21 \\
150137.22+550123.3 & 19.4 & 81.8513(3)          & CV/EC & Y & N & III,21 \\
150240.97+333423.8 & 17.6 & 84.82984(7)         & CV/EC & N & N & V,21 \\
\hline
\multicolumn{7}{l}{...continued on next page.}
\end{tabular}
\end{table*}
\begin{table*}
\textbf{Table\,1} continued.\\
\begin{tabular}{lllllll}
\hline
SDSSJ   &  $g$ & $P_\mathrm{orb}$\,[min] & Type & WD & RASS & Reference\\
\hline
150722.33+523039.8 & 18.3 & 66.612011(1)        & CV/EC & Y & N & IV,21,\tbc \\
152419.33+220920.0 & 19.1 & 94.1(1)             & DN/EC & N & N & VII,\tbc\\
153817.35+512338.0 & 18.6 & 93.11(9)            & CV    & N & N & III,13\\
154104.67+360252.9 & 19.7 & 84.3(3)             & AM    & N & N & IV,23\\
155331.11+551614.4 & 18.5 & 263.48(6)           & AM    & N & N & II,33 \\
155531.99-001055.0 & 19.4 & 113.54(3)           & CV/EC & N & N & I,3 \\
155644.23-000950.2 & 18.1 & 106.675(14)         & DN    & Y & N & I,6,17\\
155656.92+352336.6 & 18.4 & 128                 & CV/EC & N & N & V \\
160745.02+362320.7 & 18.1 & 225.36(0.63)        & NL/EC & N & N & V,13 \\
161033.63-010223.3 & 19.1 & 80.52(8)            & CV    & Y & N & I,\tbc,\tbc \\
162718.39+120435.0 & 19.2 & 150(3)*             & DN    & N & N & VII,10,\tbc \\
162936.53+263519.5 & 19.3 & 134(2)              & AM    & N & N & IV,23 \\
163722.21-001957.1 & 20.6 & 96.98(22)           & DN    & N & N & I,2 \\
164248.52+134751.4 & 18.6 & 111(2)              & CV    & N & N & VI,2 \\
165359.05+201010.4 & 18.5 & 89.7(2.0)*          & DN    & N & N & V,23\\
165658.13+212139.3 & 18.5 & 90.89(15)           & CV    & N & Y & IV,17\\
165837.70+184727.4 & 20.1 & 98.06(6)            & CV    & N & N & V,2\\
170053.29+400357.6 & 19.4 & 116.3545(1)         & AM    & N & Y & II,12 \\
170213.25+322954.1 & 17.9 & 144.118(1)          & DN/EC & N & N & III,\tbc,\tbc\\
171145.08+301319.9 & 20.3 & 80.35(5)            & CV    & Y & N & III,5 \\
173008.38+624754.7 & 16.3 & 110.22(12)          & DN    & N & N & I,37\\
204448.91-045928.7 & 16.9 & 2420(14)            & CV    & N & N & II,\tbc \\
204817.85-061044.8 & 19.4 & 87.49(32)           & CV    & Y & N & II,13,\tbc,\tbc\\
205017.83-053626.7 & 18.1 & 94.21165(3)         & AM    & N & Y & II,6,\tbc,\tbc \\
205914.87-061220.5 & 18.4 & 107.52(14)          & DN    & N & N & II,3 \\
210449.95+010545.9 & 20.4 & 103.62(12)          & DN    & N & N & II,3\\
211605.43+113407.3 & 22.5 & 80.2(2.2)           & DN    & N & N & III,5 \\
215411.12-090121.6 & 19.2 & 319(3)              & CV    & N & N & II,5 \\
210014.11+004445.9 & 18.7 & 120.8(2.0)*         & DN    & N & N & III,\tbc\\
220553.98+115553.7 & 20.1 & 82.81(9)            & CV    & Y & N & II,\tbc\\
223439.93+004127.2 & 18.1 & 127.29(25)          & CV    & N & N & II,13 \\
225831.18-094931.6 & 15.7 & 118.9(2.0)*         & DN    & N & Y & II,10 \\
230351.64+010651.0 & 19.1 & 110.51(24)          & DN    & N & Y & I,13 \\
233325.92+152222.2 & 18.7 & 83.39(8)            & IP    & N & N & V,\tbc\\
\hline                                                                  
\end{tabular}
\begin{minipage}{\textwidth}
\footnotesize
 $^{\mathrm{I}}$~\citet{szkodyetal02-2};
 $^{\mathrm{II}}$~\citet{szkodyetal03-2};
 $^{\mathrm{III}}$~\citet{szkodyetal04-1};
 $^{\mathrm{IV}}$~\citet{szkodyetal05-1};
 $^{\mathrm{V}}$~\citet{szkodyetal06-1};
 $^{\mathrm{VI}}$~\citet{szkodyetal07-2};
 $^{\mathrm{VII}}$~\citet{szkodyetal09-1};
\setcounter{tref}{1}
\tbr~Southworth et al.  in prep;      
\tbr~\citet{southworthetal08-2};      
\tbr~\citet{southworthetal07-2};      
\tbr~\citet{pretoriusetal04-1};       
\tbr~\citet{dillonetal08-1};          
\tbr~\citet{woudtetal04-1};           
\tbr~\citet{schmidtetal05-2};         
\tbr~\citet{odonoghueetal06-1};       
\tbr~\citet{southworthetal06-1};      
\tbr~\citet{katoetal09-1};            
\tbr~\citet{aketal05-1};              
\tbr~\citet{homeretal05-1};           
\tbr~Dillon et al. in prep;           
\tbr~\citet{pavlenkoetal07-1};        
\tbr~\citet{shearsetal07-1};          
\tbr~\citet{zharikovetal08-1};        
\tbr~Thorstensen et al. in prep;      
\tbr~\citet{rodriguez-giletal07-1};   
\tbr~\citet{thorstensenetal04-1};     
\tbr~\citet{boydetal08-1};            
\tbr~\citet{littlefairetal08-1};      
\tbr~\citet{woudtetal05-1};           
\tbr~this paper;                      
\tbr~\citet{schmidtetal08-1};         
\tbr~\citet{homeretal06-1};           
\tbr~\citet{schmidtetal07-1};         
\tbr~\citet{littlefairetal06-2};      
\tbr~\citet{schmidtetal05-3};         
\tbr~\citet{burleighetal06-1};        
\tbr~\citet{farihietal08-2};          
\tbr~\citet{shearsetal07-2};          
\tbr~\citet{zharikovetal06-1};        
\tbr~\citet{szkodyetal03-3};          
\tbr~\citet{wolfeetal03-1};           
\tbr~\citet{gaensickeetal06-1};       
\tbr~\citet{littlefairetal07-1};      
\tbr~Patterson et al., in prep;       
\tbr~\citet{woudt+warner04-1};        
\tbr~\citet{copperwheatetal09-1};     
\tbr~\citet{shearsetal09-1};          
\tbr~\citet{littlefairetal06-1};      
\tbr~\citet{boydetal06-2};            
\tbr~\citet{peters+thorstensen05-1};  
\tbr~\citet{woudtetal05-1};           
\tbr~\citet{woudt+warner09-1};        
\tbr~\citet{potteretal06-1};          
\tbr~\citet{homeretal06-2};           
\tbr~\citet{tramposchetal05-1};       
\tbr~\citet{southworthetal08-1};      
\tbr~\citet{southworthetal07-1}       
\end{minipage}
\end{table*}

\begin{table*}
\setcounter{tref}{1}
\label{t-prevknown}
\caption{Previously known CVs (\textit{old SDSS CVs}), which were
 spectroscopically re-identified by SDSS. The definition of the colums
are the same as in Table\,\ref{t-porbs}.}
\begin{tabular}{llllllll}
\hline
SDSSJ   & Other Name & SDSS g & $P_\mathrm{orb}$\,[min] & 
Type & WD & RASS & Reference\\
\hline
002728.01-010828.5 & EN\,Cet               & 20.7 & 85.44(7)           & DN    & Y & N & IV,\tbc  \\ 
075117.32+144423.9 & PQ\,Gem               & 14.2 & 311.56(4)          & IP    & N & Y & V,\tbc   \\
075853.03+161645.1 & DW\,Cnc   	           & 15.3 & 86.1015(3)         & IP    & N & N & V,\tbc,\tbc \\
082236.03+510524.5 & BH\,Lyn               & 15.3 & 224.460831(20)     & NL/EC & N & N & I,\tbc,\tbc,\tbc,\tbc \\
083619.15+212105.3 & CC\,Cnc               & 16.8 & 105.86(7)          & DN    & N & Y & V,\tbc \\
083642.74+532838.0 & SW\,UMa               & 16.9 & 81.8136(1)         & DN    & N & Y & I,\tbc,\tbc \\
084303.98+275149.6 & EG\,Cnc               & 18.9 & 85.5(9)            & DN    & Y & N & IV,\tbc,\tbc \\
085107.39+030834.3 & CT\,Hya               & 18.8 & 93.9(2.0)*         & DN    & N & N & II,\tbc,\tbc\\
085344.16+574840.5 & BZ\,UMa               & 16.4 & 97.8(1)            & DN    & N & Y & II,\tbc \\
085414.02+390537.2 & EUVE\,J0854+390       & 19.2 & 113.26(3)          & AM    & N & Y & IV,1 \\
085909.18+053654.5 & RX\,J0859.1+0537      & 18.6 & 143.8              & AM    & N & Y & IV,\tbc \\
090950.53+184947.4 & GY\,Cnc               & 16.4 & 252.6371983(30)    & DN/EC & N & Y & VII,\tbc,\tbc \\
093214.82+495054.7 & 1H\,0928+5004         & 17.5 & 602.45813(43)      & NL/EC & N & N & V,\tbc  \\
093836.98+071455.0 & PG\,0935+075          & 18.3 & 269.0(4)           & DN    & N & N & IV,\tbc \\
094431.71+035805.5 & RXJ\,0944.5+0357      & 16.8 & 214.4(2)           & DN    & N & Y & II,19,\tbc,\tbc\\
094636.59+444644.7 & DV\,UMa               & 19.4 & 123.6278190(20)    & DN/EC & N & N & IV,\tbc,\tbc,\tbc \\
101534.67+090442.0 & GG\,Leo               & 17.2 & 79.879464(66)      & AM    & N & Y & IV,\tbc \\
101947.26+335753.6 & HS\,1016+3412         & 18.4 & 92.22(17)          & DN    & N & Y & VI,\tbc\\
102026.52+530433.1 & KS\,UMa               & 17.4 & 97.86(14)          & DN    & N & Y & III,22,\tbc \\
102320.27+440509.8 & NSV\,4838, UMa\,8     & 18.8 & 97.8(3)            & DN    & N & N & IV,20 \\
102800.07+214813.5 & 1H\,1025+220          & 16.0 & 210.36             & EC    & N & N & VII,\tbc\\
104356.72+580731.9 & IY\,UMa               & 17.7 & 106.42892(7)       & DN/EC & N & N & VII,\tbc,\tbc,\tbc\\
105135.14+540436.0 & EK\,UMa               & 18.4 & 114.5(2)           & AM    & N & N & IV,\tbc \\
105430.43+300610.1 & SX\,LMi               & 16.8 & 96.72(16)          & DN    & N & Y & VI,\tbc,\tbc \\
105656.99+494118.2 & CY\,UMa               & 17.8 & 100.18(6)          & DN    & N & N & IV,\tbc,\tbc \\
110425.64+450314.0 & AN\,UMa               & 15.8 & 114.84406(6)       & AM    & N & N & V,\tbc,\tbc \\
110539.76+250628.6 & ST\,LMi               & 17.6 & 113.8882(1)        & AM    & Y & Y & VII,\tbc\\
111544.56+425822.4 & AR\,UMa               & 15.6 & 115.92107(17)      & AM    & N & N & V,\tbc,\tbc \\
113122.39+432238.5 & RX\,J1131.3+4322, MR\,UMa & 16.2 & 91.25(12)      & DN    & N & N & V,\tbc \\
113722.24+014858.5 & RZ\,Leo               & 18.7 & 109.6(2)           & DN    & Y & Y & II,29 \\
113826.82+032207.1 & T\,Leo                & 14.9 & 84.6994(7)         & DN    & N & Y & II,\tbc \\
114955.69+284507.3 & EU\,UMa               & 17.9 & 90.14(2)           & AM    & N & Y & VII,\tbc\\
115215.82+491441.8 & BC\,UMa               & 18.5 & 90.16(6)           & DN    & Y & N & III,29 \\
125637.10+263643.2 & GO\,Com               & 18.3 & 95(1)              & DN    & N & Y & VII,\tbc,\tbc,\tbc\\
130753.86+535130.5 & EV\,UMa               & 16.5 & 79.687973(29)      & AM    & N & Y & IV,\tbc,\tbc \\
134323.16+150916.8 & HS\,1340+1524         & 17.6 & 92.66(17)          & DN    & N & Y & VII,28\\
143500.21-004606.3 & OU\,Vir               & 18.6 & 104.696803(7)      & DN/EC & N & N & I,\tbc,\tbc\\
151302.29+231508.4 & NY\,Ser               & 16.4 & 140.4(3)           & DN    & N & N & VII,29,\tbc \\
152613.96+081802.3 & QW\,Ser               & 18.1 & 107.3(1)           & DN    & N & N & VII,29 \\

155247.18+185629.1 & MR\,Ser               & 17.2 & 113.4689(1)        & AM    & N & Y & VII,\tbc \\
155412.33+272152.4 & RX\,J1554.2+2721      & 17.6 & 151.865(9)         & AM    & N & Y & VI,\tbc,\tbc\\
155654.47+210718.8 & QZ\,Ser               & 17.9 & 119.752(2)         & DN    & N & N & VI,\tbc\\
161007.50+035232.7 & RX\,J1610.1+0352	   & 17.7 & 190.54(6)          & AM    & N & Y & VII,\tbc,\tbc\\
162501.74+390926.3 & V844\,Her             & 17.2 & 78.69(1)           & DN    & N & Y & IV,\tbc  \\     
223843.84+010820.7 & Aqr\,1                & 18.3 & 194.30(16)         & IP    & N & N & II,\tbc,\tbc \\
\hline                                                                  
\end{tabular}
\setcounter{tref}{1}
\begin{minipage}{\textwidth}
 $^{\mathrm{I}}$~\citet{szkodyetal02-2};
 $^{\mathrm{II}}$~\citet{szkodyetal03-2};
 $^{\mathrm{III}}$~\citet{szkodyetal04-1};
 $^{\mathrm{IV}}$~\citet{szkodyetal05-1};
 $^{\mathrm{V}}$~\citet{szkodyetal06-1};
 $^{\mathrm{VI}}$~\citet{szkodyetal07-2};
 $^{\mathrm{VII}}$~\citet{szkodyetal09-1};
\tbr~\citet{dillonetal08-1};          
\tbr~\citet{hellieretal94-2};         
\tbr~\citet{rodriguez-giletal04-1};   
\tbr~\citet{pattersonetal04-1};       
\tbr~\citet{thorstensenetal91-2};     
\tbr~\citet{dhillonetal92-1};         
\tbr~\citet{hoard+szkody97-1};        
\tbr~\citet{stanishevetal06-1};       
\tbr~\citet{thorstensen97-1};         
\tbr~\citet{howell+szkody88-1};       
\tbr~\citet{shafteretal86-2};         
\tbr~\citet{pattersonetal98-4};       
\tbr~\citet{katoetal04-1};            
\tbr~\citet{nogamietal96-1};          
\tbr~\citet{katoetal99-3};            
\tbr~\citet{ringwaldetal94-1};        
\tbr~Reinsch, priv. comm.;            
\tbr~\citet{gaensickeetal00-2};       
\tbr~\citet{felineetal05-1};          
\tbr~Thorstensen, in prep..;          
\tbr~\citet{thorstensen+taylor01-1};  
\tbr~\citet{jiangetal00-1};           
\tbr~\citet{mennickentetal02-1};      
\tbr~\citet{howelletal88-1};          
\tbr~\citet{pattersonetal00-2};       
\tbr~\citet{felineetal04-2};          
\tbr~\citet{burwitzetal98-1};         
\tbr~\citet{aungwerojwitetal06-1};    
\tbr~\citet{pattersonetal03-1};       
\tbr~\citet{taylor99-1};              
\tbr~\citet{uemuraetal00-2};          
\tbr~\citet{pattersonetal00-3};       
\tbr~\citet{steeghsetal03-1};         
\tbr~\citet{morrisetal87-1};          
\tbr~\citet{nogamietal97-1};          
\tbr~\citet{wagneretal98-1};          
\tbr~\citet{martinez-pais+casares95-1}; 
\tbr~\citet{thorstensenetal96-1};     
\tbr~\citet{liebertetal82-1};         
\tbr~\citet{bonnet-bidaudetal96-1};   
\tbr~\citet{cropper86-1};             
\tbr~\citet{remillardetal94-2};       
\tbr~\citet{schmidtetal99-2};         
\tbr~\citet{pattersonetal05-3};       
\tbr~\citet{shafter+szkody84-1};      
\tbr~\citet{howelletal95-2};          
\tbr~\citet{howelletal95-3};          
\tbr~\citet{howelletal90-1};          
\tbr~Howell, priv. comm.;             
\tbr~\citet{osborneetal94-1};         
\tbr~\citet{katajainenetal00-1};      
\tbr~\citet{vanmunsteretal00-1};      
\tbr~\citet{felineetal04-1};          
\tbr~\citet{nogamietal98-1};          
\tbr~\citet{schwopeetal91-1};         
\tbr~\citet{tovmassianetal01-2};      
\tbr~\citet{thorstensen+fenton02-1};  
\tbr~\citet{thorstensenetal02-2};     
\tbr~\citet{schwopeetal02-2};         
\tbr~\citet{rodriguesetal06-1};       
\tbr~\citet{thorstensenetal02-3};     
\tbr~\citet{woudtetal04-1};           
\tbr~\citet{bergetal92-1};            
\tbr~\citet{southworthetal08-2}       
\end{minipage}
\end{table*}

Even if the nature of the variability used to determine the orbital
period is unambiguous, the details of the temporal sampling of the
observations have a crucial impact on the quality of the orbital
period measurement that can be obtained from the observational
data. In particular, the accuracy of the period measurement is limited
by the length of the observations. Except for co-ordinated
multi-longitude observations, the immediate limit on this is set by
the seasonal visibility of a given object, which will in the best case
be equal to the length of the observing night. Given that CV periods
are in the range from 80\,min to about a day, a single night of data
will hence cover at most a few orbital cycles, and the typical lower
limit on the error of orbital periods from such data is
$\sim10$\%. More accurate period determinations are obtained by
combining data obtained over a number of nights. However, this
potential for an improved period measurement comes at the price of an
additional uncertainty. As the CV orbital periods are often much
shorter than a day, the number of orbital cycles in between the two, or
more, observing nights may be ambiguous, which is known as cycle-count
uncertainty, or aliasing.  \citet{thorstensen+freed85} discuss in
detail Monte-Carlo techniques that allow the probability of different
cycle-count aliases to be established. A different approach using
bootstrapping is given by \citet{southworthetal06-1}.

\begin{figure}
 \includegraphics[width=\columnwidth]{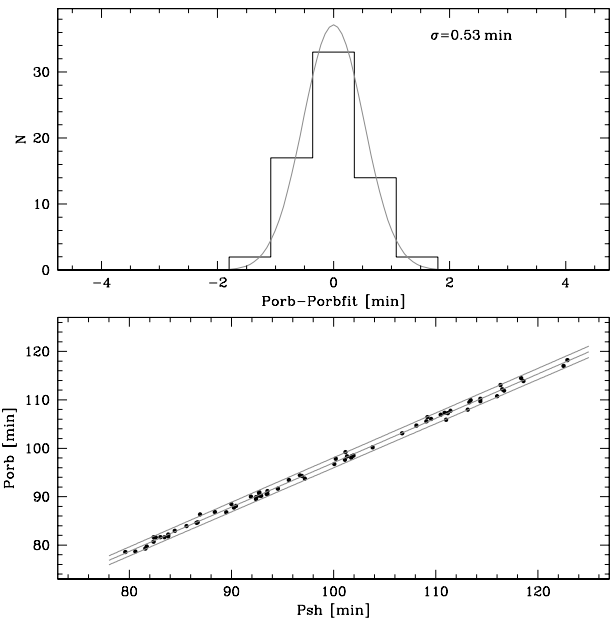}
  \caption{\label{f-psh-porb} The $\Psh-\Porb$
      relation. Bottom panel: 68 systems from
      \citet{pattersonetal05-3} with $\Porb<120$\,min, i.e. below the
      period gap. Shown as gray lines are the best linear fit and its
      $1\,\sigma$ errors. Top panel: the histogram shows the
      distribution of observed minus computed (Eq.\,\ref{e-psh-porb})
      orbital periods. The distribution is well-fit by a Gaussian with
      a standard deviation of $0.53$\,min.}
\end{figure}

Our aim in this paper is to establish the detailed orbital period
distribution of the SDSS CVs\footnote{We exclude from our analysis the
  AM\,CVn stars, which are ultrashort-period helium-rich
  double-degenerate binaries that are not included in the standard CV
  population models. We also exclude SDSSJ102347.68+003841.1
  (FIRSTJ1023.6+003841, \citealt{bondetal02-1,thorstensen+eve05-1,
    homeretal06-1}) which has recently been confirmed as a neutron
  star binary.}, which represent the largest CV sample selected
in a homogenous way. For this purpose, we inspected the published
orbital period measurements of every individual SDSS CV according to
the caveats mentioned above, and excluded systems where the relative
error on the orbital period measurement exceeded 3\%, or where we had
substantial doubt about the correct choice of the cycle count alias.

In addition to direct orbital period measurements, we also considered
indirect orbital period estimates for CVs exhibiting photometric
variability in the form of superhumps, which are thought to be related
to tidal interactions between the donor star and the accretion
disc. \citet{stolz+schoembs84-1} published an empirical relation
between the superhump period (\Psh) and the orbital period, which has
been updated on several occasions throughout the years. We decided to
investigate the method once more. \citet{pattersonetal05-3} provides a
recent compilation of CVs with reliable orbital and superhump
periods. While there is a clear correlation of the superhump excess
$\epsilon=(\Psh-\Porb)/\Porb$ and the orbital period, a few clear
outliers are present at short orbital periods and small values of
$\epsilon$, most of which are ``WZ\,Sge'' systems (CVs with very long
intervals between dwarf nova outbursts and extreme mass ratios). We
restricted the following analysis to CVs below the period gap, where
the bulk of the well-studied (and well-behaved) systems lie, and where
the majority of the application of this method will take place. This
leaves 68 CVs from the list of \citet{pattersonetal05-3}
(Fig.\,\ref{f-psh-porb}, bottom panel). We then fitted a linear
relation to the ($\Psh,\Porb$) data,
\begin{equation}
\Porb=0.9192(52)\Psh+5.39(52)
\label{e-psh-porb}
\end{equation}
where both periods are given in minutes. In order to assess the error
of orbital periods estimated in this way, we calculated for all 70
systems \Porb\ values from Eq.\,\ref{e-psh-porb} and computed the
differences from their actual orbital periods. The observed minus
calculated \Porb{}s follow approximately a Gaussian distribution with
a standard deviation of $0.53$\,min (Fig.\,\ref{f-psh-porb}, top
panel). We conclude that accurate estimates of orbital periods can be
obtained from reliable superhump periods.  In the context of this
paper, we assume a somewhat more pessimistic error on the
superhump-based orbital periods of 2\,min, which accounts for the small
drifts in \Psh\ often observed during the evolution of the dwarf nova
outburst. The orbital periods obtained from superhumps included in our
analysis below range from 89.6\,min (SDSS\,J1653+2010) to 150\,min
(SDSS\,J1627+1204), with the latter one being the only system for
which we extrapolated Eq.\,(\ref{e-psh-porb}) above the range used for
the calibration of the method.

The SDSS CVs that passed our scrutiny are listed in
Table\,\ref{t-porbs} for CVs that were identified from SDSS
spectroscopy, and in Table\,\ref{t-prevknown} for those CVs which were
previously known, and re-identified as CVs from the SDSS spectra.

\section{The orbital period distribution of the SDSS cataclysmic variables}
\label{s-period_dist}

The fibre allocation of SDSS does not cross-correlate with
astronomical catalogues such as Simbad, and hence the nature of
targets for spectroscopic follow-up may be known prior to the SDSS
observation. For the following discussion, we shall call \textit{SDSS
  CVs} all CVs for which an SDSS spectrum that allows their
identification as a CV is available in Data Release 6 (DR6,
\citealt{adelman-mccarthyetal07-1}), independently of whether they
were already known before or not. Furthermore, we shall call
\textit{new SDSS CVs} those systems that were genuinely identified
from SDSS spectroscopy, and \textit{old SDSS CVs} those systems which
were known to be CVs beforehand, and were found again among the SDSS
spectra. Finally, we shall call \textit{non-SDSS CVs} all CVs from
V7.6 of the Ritter \& Kolb \citeyear{ritter+kolb03-1} catalogue which
have \textit{no} spectrum in SDSS~DR6, and excluding the systems which
were flagged as having an uncertain orbital period measurement. The
four different samples contain 454 (non-SDSS CVs), 137 (SDSS CVs), 92
(new SDSS CVs) and 45 (old SDSS CVs) systems.

\begin{figure*}
 \includegraphics[width=\columnwidth]{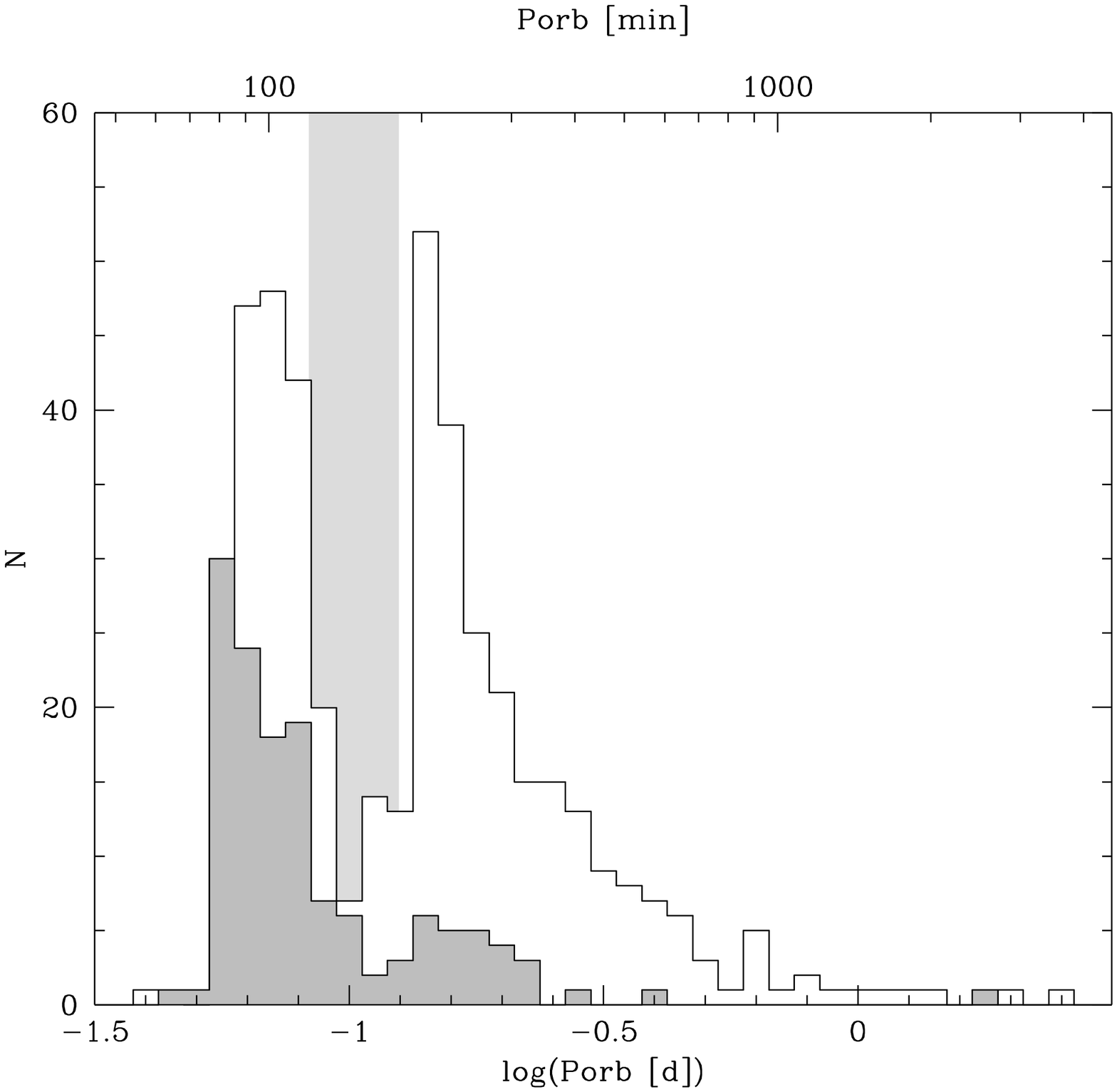}
 \includegraphics[width=\columnwidth]{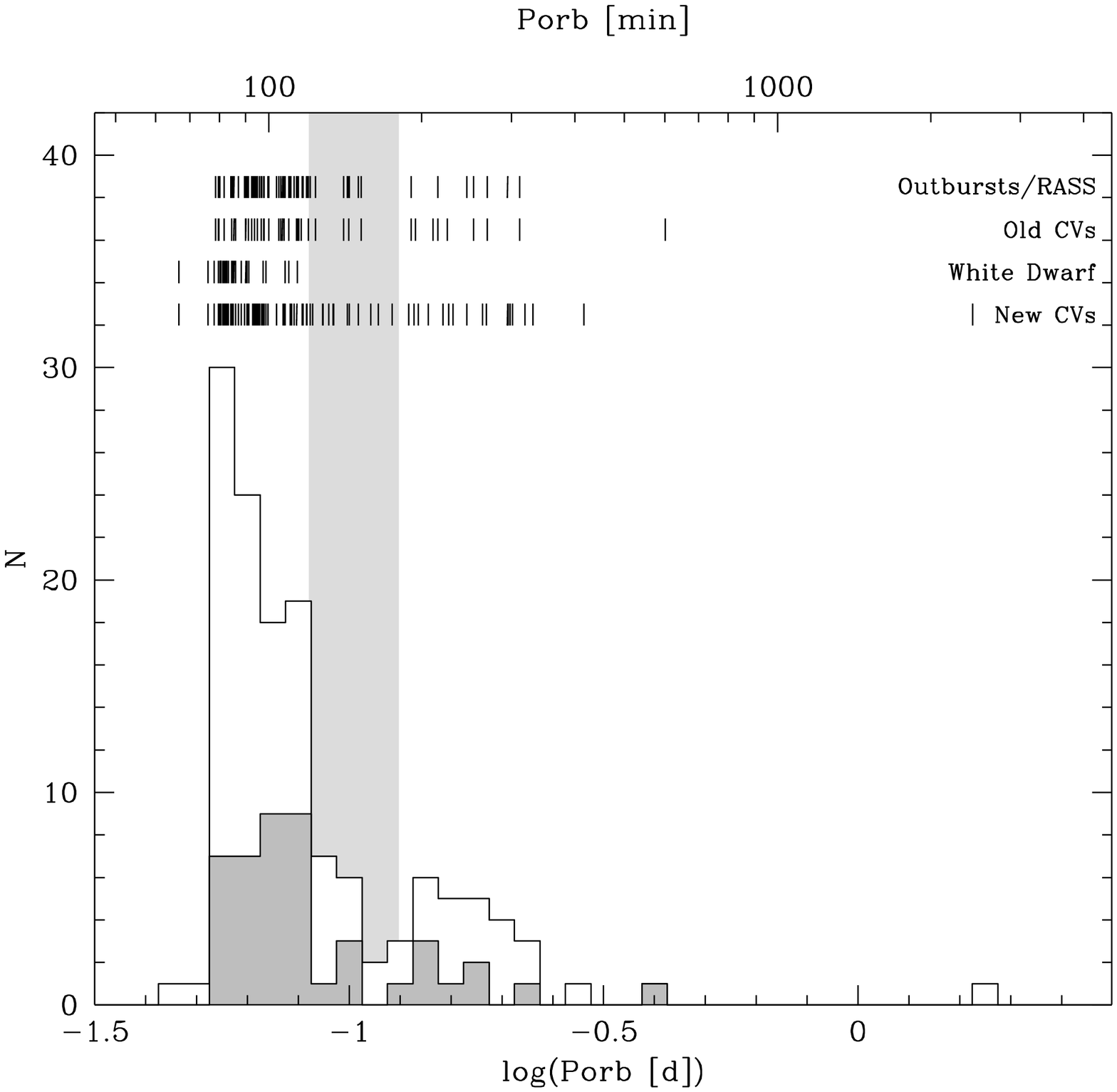}
 \caption{\label{f-porbdist} Left panel: The orbital period
   distribution of 454 CVs from \citet{ritter+kolb03-1}, V7.6, which
   have no spectroscopic observation in SDSS DR6 (white) and the
   distribution of 137 SDSS CVs from Table\,\ref{t-porbs} \&
   \,\ref{t-prevknown} (\textit{SDSS CVs}, gray). The gray shaded area
   represents the 2--3\,h orbital period gap. These distributions
   exclude the hydrogen-deficient AM\,CVn systems. Right panel: the
   period distribution of the SDSS CVs divided into 45 previously
   known systems (\textit{old SDSS CVs}, gray) and 92 newly identified
   CVs (\textit{new SDSS CVs}, white). Superimposed are tick marks
   indicating the individual orbital periods of the old and new SDSS
   CVs, along with the periods of SDSS CVs showing outbursts and/or
   being detected in the ROSAT All Sky Survey, and the distribution of
   the periods of the SDSS CVs which clearly reveal the white dwarf in
   their optical spectra. The bin width in both panels is
   $0.05\log(\mathrm{d})$. }
\end{figure*}

Figure\,\ref{f-porbdist} compares the orbital period distribution of
454 non-SDSS CVs and that of 137 SDSS CVs. The well-known features of
the non-SDSS CV population are the deficiency of systems in the
2--3\,h period gap, roughly equal numbers of systems above and below
the gap, a minimum period near 80\,min, and a drop-off of systems towards
longer periods above the gap.  In numbers, the non-SDSS CV sample
contains 167, 48, and 239 systems with $\Porb\le2$\,h,
$2\,\mathrm{h}<\Porb\le3$\,h, and $\Porb>3$\,h, respectively.

The period distribution of the SDSS CVs looks radically different
compared to that of the non-SDSS CVs. The majority of the systems are
found below the period gap, with 92, 17, and 29 systems below, in, and
above the 2--3\,h period gap, respectively, confirming the trend
already noticed by \citet{szkodyetal03-2, szkodyetal07-2} and
\citet{southworthetal06-1, southworthetal07-2}. The most striking
feature is an apparent accumulation of systems in the shortest period
bin. Comparing the orbital period distributions of the old SDSS CVs
and the new SDSS CVs sample (Fig.\,\ref{f-porbdist}, right panel)
indicates that the distribution of the old SDSS CVs is flat below the
period gap, and that the ``spike'' at the orbital period minimum comes
entirely from the new CVs identified by SDSS.

\begin{figure}
 \includegraphics[angle=-90,width=\columnwidth]{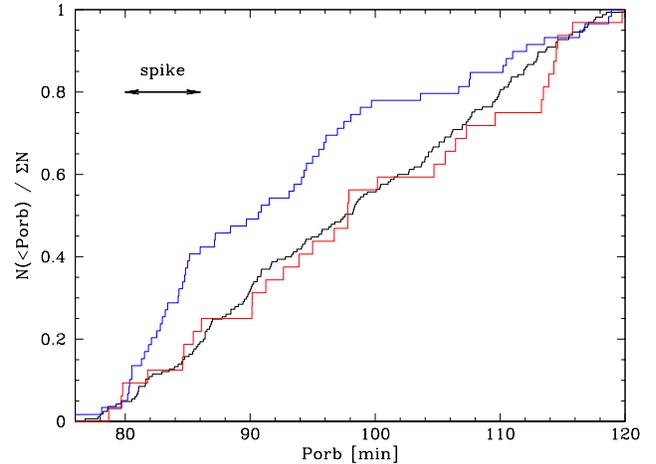}
 \caption{\label{f-cumulative} Normalised cumulative period
   distributions in the range $\Porb=76-120$\,min of (a) 454 CVs from
   \citeauthor{ritter+kolb03-1} (2003; V7.6), excluding systems with
   spectroscopy in SDSS DR6 and systems whose periods are marked as
   uncertain (black), (b) 45 previously known CVs recovered from SDSS
   spectroscopy (red), and (c) 92 new CVs identified from SDSS
   spectroscopy (blue). A two-sided Kolmogorov-Smirnov test comparing
   the distributions of the non-SDSS CVs and the new SDSS CVs results
   in a $9.5\times10^{-3}$ probability for the two distributions being
   drawn from the same parent population. In contrast, the probability
   that the non-SDSS CVs and the previously known SDSS CVs are drawn
   from the same sample is 57.4\%.}
\end{figure}

Cumulative period distributions of the non-SDSS CVs, the old SDSS CVs,
and the new SDSS CVs are shown on a linear scale in orbital period in
Fig.\,\ref{f-cumulative}.  The large number of new CVs among the new
SDSS CVs near the period minimum is reflected in the rapid rise of the
cumulative distribution over the range $\sim80-86$\,min, with a clear
break in slope at $\sim86$\,min. In Sect.\,~\ref{s-spike_prop}, we
will inspect the properties of the systems in this $\sim80-86$\,min
period spike.  

We have applied a two-sided Kolmogorov-Smirnov (KS) test in order to
test whether the three cumulative distributions deviate from one
another in a statistically significant way.  As a lower limit, we chose
76.78\,min, corresponding to the shortest-period ``standard''
hydrogen-rich CV, GW\,Lib\footnote{Three hydrogen-rich CVs with
  orbital periods around 60\,min are known: V485\,Cen, EI\,Psc, and
  SDSS\,J1507+5230. The evolutionary state of these systems is not
  entirely clear; EI\,Psc contains an evolved donor star
  \citep{thorstensenetal02-1}, and SDSS\,J1507+5230 may have formed
  with a brown dwarf donor \citep{littlefairetal07-1} or be a halo CV
  \citep{pattersonetal08-1}. For the the discussion of the SDSS CV
  period distribution, we decided to exclude these three oddball
  systems.}. We set the upper period limit for the KS test to
120\,min, i.e. the lower edge of the period gap. The reason for this
choice is that CVs with periods above the gap have substantially
fainter absolute magnitudes compared to the short-period systems, and
hence SDSS, sampling the sky at high galactic latitudes,
$|b|>30^\circ$, will be biased against the detection of long-period
CVs\footnote{Examples of luminous CVs above the period gap are
  TT\,Ari, with $\Porb=198.07$\,min, $V\simeq10.6$, and
  $d=335\pm50$\,pc \citep{thorstensenetal85-1, gaensickeetal99-1}, and
  MV\,Lyr with $\Porb=191.4$\,min, $V\simeq12.0$, and $d=505\pm50$\,pc
  \citep{skillmanetal95-1, hoardetal04-1}, corresponding to absolute
  magnitudes of $M_V=3.0\pm0.3$ and $M_V=3.5\pm0.2$, respectively. At
  $|b|=30^{\circ}$, SDSS could detect these systems out to $>10$\,kpc
  above the galactic disc. See Sect.\,\ref{s-deep} for details on the
  sky volume coverage of SDSS.}. In fact, a substantial fraction of
previously known long-period CVs contained in the SDSS footprint are
saturated in the SDSS imaging data, were consequently not selected by
the SDSS target algorithms for spectroscopic follow-up, and are hence
not available in the SDSS CV sample.

The KS test comparing the cumulative period distributions of the
non-SDSS CVs and the new SDSS CVs results in a $9.5\times10^{-3}$
probability that the two distributions are randomly drawn from an
identical parent population, which shows that the period distribution
of the new SDSS CVs differs significantly from that of the previously
known CV sample. This clearly suggests that the CV identification
within SDSS differs from the average CV discovery method in the
previously known sample.

Conversely, comparing the cumulative period distributions of the
non-SDSS CVs and the old SDSS CVs, the probability for both
distributions emanating from the same parent sample is 57.4\%,
i.e. the two distributions are identical from a statistical point of
view. This is not surprising, as the old CVs have been identified
by the same methods as the non-SDSS CVs, i.e. primarily dwarf nova
outbursts and X-ray emission. 

We conclude for now that the orbital period distributions of the new
SDSS CVs and of the previously known CVs (independently of whether
or not they were also selected by SDSS for spectroscopic follow-up) differ
significantly, most apparently in the ratio of the number of short and
long period systems, and in the appearance of a spike at
$\sim80-86$\,min in the period distribution of the new SDSS CVs. In
the next Section, we will discuss in more detail additional
differences between the new SDSS CV and non-SDSS CV samples.

\section{Properties of the period-spike cataclysmic variables found by SDSS}
\label{s-spike_prop}

As shown in Sect.\,\ref{s-period_dist}, the orbital period
distributions of the non-SDSS CVs and the new SDSS CVs differ at a
$>3\,\sigma$ level. This raises the question of whether the new SDSS
CVs also differ in other properties besides their orbital period from
the non-SDSS CVs. For the discussion below, we have defined the
following three boolean characteristics: (a) the white dwarf is
clearly visible in the SDSS identification spectrum, (b) X-ray
emission has been detected in the ROSAT All Sky Survey (RASS,
\citealt{vogesetal00-1})\footnote{While a number of SDSS CVs have been
  detected in pointed ROSAT observations, we restricted our assessment
  of X-ray emission to a detection in the RASS, as inclusion of
  pointed observations would imply wildly different X-ray flux limits
  for random lines-of-sight.}, and (c) an optical outburst has been
observed, leading to the classification as a dwarf
nova\footnote{Information on large-amplitude ($>1$\,mag) variability
  has been drawn from three sources of information. (1) Brightness
  differences between the SDSS imaging and spectroscopic data, (2)
  individual follow-up observations which caught some systems in
  outburst \citep[e.g.][]{tramposchetal05-1, dillonetal08-1,
    southworthetal07-2}, and from the mailing lists of the amateur
  astronomers (vsnet, cvnet, baavss). We do \textit{not} include
  outburst information from robotic telescopes, such as e.g. the Catalina
  Real-Time Transient Survey \citep{drakeetal09-1}, \textit{unless}
  they were announced via one of the appropriate mailing
  lists.}. Obviously, the non-detection of X-rays or outbursts is only
an average characteristic, as the observations of an individual system
may have just missed such activity. Tables\,~\ref{t-porbs} and
\ref{t-prevknown} list these three properties for the 137 SDSS CVs
with accurate orbital period measurements. Below, we will interpret
the detection of the white dwarf in the optical spectrum as evidence
for a low mass transfer rate, and the detection of X-ray emission
and/or outbursts as evidence of accretion activity which may lead to
the identification of a system as a CV.

\begin{figure}
 \includegraphics[width=\columnwidth]{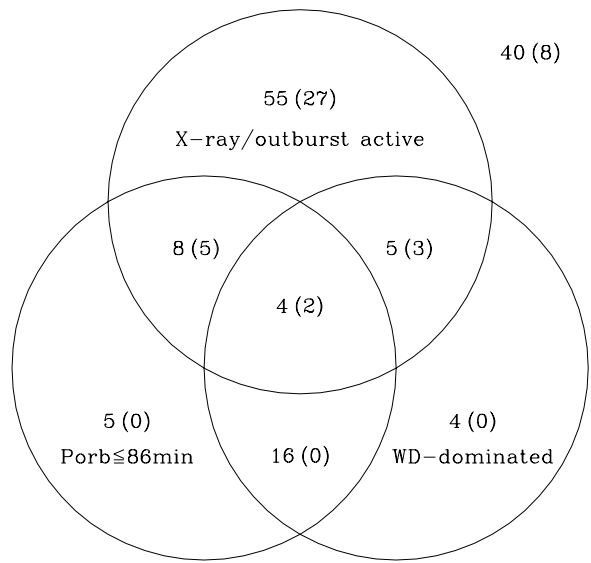}
 \caption{\label{f-venn} Venn diagram showing the distribution of 85
   SDSS CVs with accurate orbital periods (Tables\,~\ref{t-porbs} \&
   \ref{t-prevknown}) into the following three categories:
   "$\Porb\le86$\,min", "has outbursts and/or has been detected in the
   ROSAT All Sky Survey", and "its optical spectrum is dominated by
   the white dwarf". The numbers in brackets refer to the properties
   of the previously known CVs that were re-identified by SDSS. Not
   all the systems from Tables\,\ref{t-porbs} are represented in this
   diagram, 40 additional SDSS CVs from Tables\,~\ref{t-porbs} \&
   \ref{t-prevknown} have $\Porb>86\,$min, do not exhibit the white
   dwarf in their SDSS spectra, and neither been detected in the RASS,
   nor seen in outburst.}
\end{figure}

In Fig.\,\ref{f-venn}, we visualise the properties of the SDSS CVs in
the form of a three-set Venn-diagram, dividing them into systems with
(1) $\Porb\le86$\,min (i.e. within the period spike), (2) systems
which reveal the white dwarf in the SDSS identification spectrum, and
(3) systems which were detected in X-rays and/or outburst. The numbers
for the previously known systems are given in parentheses. An
immediate result from inspection of Fig.\,\ref{f-venn} is that all the
previously known systems show accretion activity, whereas 25 of the
new SDSS CVs have not been detected in the RASS and have not (yet)
been detected in outburst so far.

Figure\,\ref{f-venn} clearly illustrates that a large fraction (20/33)
of the systems with orbital periods in the 80--86\,min spike have
optical spectra dominated by the white dwarf. This generally indicates
that the accretion disc and secondary star are dim and hence that the
donor star is of a late spectral type and of low mass, and that the
mass transfer rate is low. Of the systems not revealing their white
dwarfs, eight exhibit accretion activity, which is (on average)
indicative of somewhat higher mass transfer rates than in the
white-dwarf dominated systems, and result in the white dwarf being
outshone by the accretion disc/stream. In fact, five of these eight
systems were discovered because of their outbursts or X-ray emission
(SW\,UMa, GG\,Leo, T\,Leo, EV\,UMa, and V844\,Her). The remaining
three are the polar SDSS\,J1541+2721, and the dwarf novae
SDSS\,J1250+6655 and SDSS\,J2116+1134. In magnetic CVs, the white
dwarf is typically not detected during states of active accretion,
SDSS\,J1250+6655 is an eclipsing system, where the white dwarf may be
obstructed by the accretion disc, and the spectrum of SDSS\,J2116+1134
is too poor to make a definite judgement on the presence of broad
white dwarf absorption lines.

\begin{figure}
 \includegraphics[width=\columnwidth]{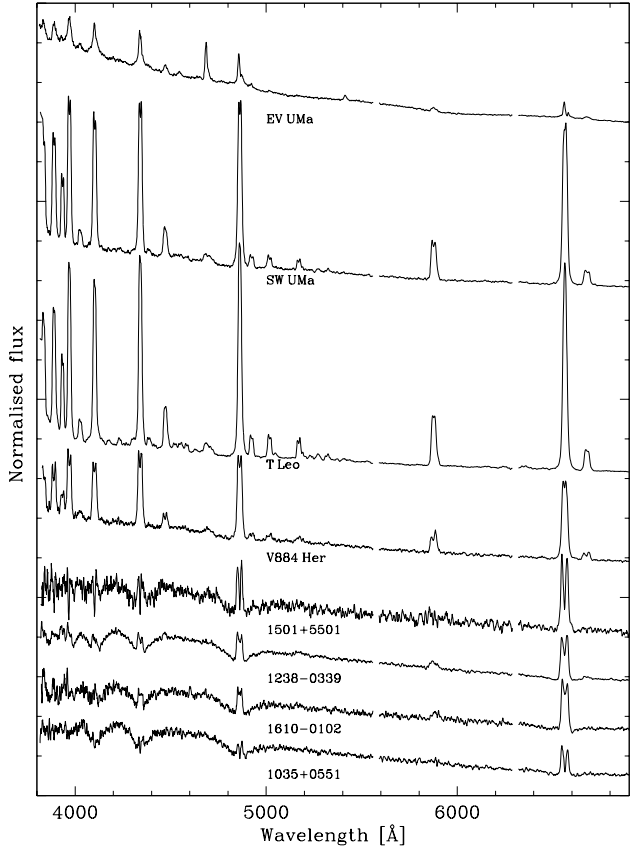}
 \caption{\label{f-compspec} SDSS CVs with periods below 86\,min. The
   top four objects are previously known systems re-identified by SDSS
   (\textit{old SDSS CVs}, Table\,\ref{t-prevknown}), the bottom four
   spectra are new identifications (\textit{new SDSS CVs},
   Table\,\ref{t-porbs}).}
\end{figure}

The very distinct spectral appearance between the CVs with
$\Porb\le86$\,min which are accretion-active and those which are not
is demonstrated in Fig.\,~\ref{f-compspec}, where four representative
spectra from each group are shown. Clearly, the four old
(accretion-active) SDSS CVs EV\,UMa, SW\,UMa, T\,Leo, and V884\,Her
have strong Balmer and \Ion{He}{I} emission lines, or \Ion{He}{II} in
the case of the polar EV\,UMa. These four systems have been identified
as an X-ray emitter (EV\,UMa) and those outbursting dwarf novae (SW\,UMa,
T\,Leo, V884\,Her).  The four white-dwarf dominated new SDSS CVs,
SDSS\,J1501+5501, SDSS\,J1238--0339, SDSS\,J1610--0102, and
SDSS\,J1035+0551 have moderate to weak Balmer emission lines with a
steep decrement, and hardly any \Ion{He}{I} emission, indicating lower
temperatures and emission measures in their accretion flows than in
the accretion-active CVs. SDSS\,J1501+5501 and SDSS\,J1035+0551 have
dwarf donor stars with sub-stellar masses \citep{littlefairetal06-2,
  littlefairetal08-1}, indicating that they are probably highly
evolved CVs, and the white dwarf in SDSS\,J1610--0102 exhibits
ZZ\,Ceti pulsations \citep{woudt+warner04-1}, which implies a
relatively cool white dwarf and a low secular mean accretion rate
\citep{townsley+bildsten03-1, arrasetal06-1}.

In addition to the eight accretion-active CVs with $\Porb\le86$\,min
where the white dwarf cannot be detected in their SDSS spectra, there
are another five systems that fail to reveal the white dwarf and have
also not been detected in X-rays and/or outburst. Two of these systems
are eclipsing (SDSS\,J0903+3300, and SDSS\,J1502+3334). Given that the
strength of emission lines is positively correlated with inclination
\citep{warner86-1}, the white dwarf Balmer absorption lines are likely
to be filled in by emission from the optically thin accretion flow
and/or the white dwarf might be veiled by the accretion disc
altogether \citep{horneetal94-1, kniggeetal00-1}. The other three
systems are magnetic CVs (SDSS\,J1031+2028, SDSS\,J0921+2038, and
SDSS\,J2333+1522), where the white dwarf is typically not seen during
accretion-active phases.

Of particular interest are the four systems with $\Porb\le86$\,min
which are accretion-active, and do reveal the white dwarf in their
optical spectra: the two previously known systems EG\,Cnc (a
well-studied WZ\,Sge-type dwarf nova) and EN\,Cet (a poorly studied
dwarf nova), and the two new SDSS dwarf novae SDSS\,J0137-0912 (with a
single observed superoutburst) and SDSS\,J0804+5103 (a WZ\,Sge dwarf
novae with the first outburst observed in March 2005). All four
systems are characterised by long outburst recurrence times, which is
a signature of CVs with low mass transfer rates, and consistent with
the detection of their white dwarfs in the SDSS spectra.

A final note on Fig.\,\ref{f-venn} is that only 9 out of 104 SDSS CVs
with $\Porb>86$\,min have white-dwarf dominated optical spectra,
confirming that the detection of the white dwarf is indeed a
spectroscopic fingerprint of the shortest period CVs.

Two additional properties of SDSS CVs with periods below
  86\,min are worth mentioning, as they again suggest a predominance of low
  accretion rates and low-mass companion stars near the period
  minimum: (1) A number of non-radial white dwarf pulsators have been
identified by SDSS. For six of them accurate orbital periods have been
measured, and with the exception of the ultra-short period system
SDSS\,J1507+5230 \citep{pattersonetal08-1}, they all reside within the
80--86\,min period spike \citep{woudt+warner04-1, woudtetal05-1,
  gaensickeetal06-1, nilssonetal06-1, mukadametal07-1}. If a
sufficient number of pulsation mode frequencies can be identified, it
will be possible to measure the white dwarf core and envelope masses
in these systems \citep{townsley+bildsten03-1,
  townsley+gaensicke09-1}.  Obviously, the identification of white
dwarf pulsations relies in the first place on the clear detection of
the white dwarf in the optical wavelength range. In contrast to single
white dwarfs, where pulsators occupy a well-defined instability strip
in the ($\log g,\Teff$) plane \citep{mukadametal04-2,
  gianninasetal06-1}, the temperatures of the pulsators in CVs span a
relatively large range in effective temperature \citep{szkodyetal02-4,
  araujo-betancoretal05-1, szkodyetal07-1}. Theoretical work suggests
that the presence of helium in the accreted material may result in
additional driving mechanisms \citep{arrasetal06-1}. (2) Three of the
four SDSS CVs with confirmed donor stars with sub-stellar masses are
located within the period spike (SDSS\,J1035+0551, SDSS\,J1433+1011,
and SDSS\,J1501+5501, \citealt{littlefairetal06-2,
  littlefairetal08-1}); \footnote{The fourth confirmed brown dwarf
  donor has been found in SDSS\,J1507+5230 \citep{littlefairetal07-1,
    pattersonetal08-1}, which has been excluded from the discussion in
  this paper as it has an orbital period of 66.6\,min, way below the
  period minimum for ``standard'' hydrogen-rich CVs.}, these are
probably CVs that have evolved past the period minimum.

In summary, twenty of the 33 SDSS CVs in the 80--86\,min period spike
are systems that differ dramatically from the bulk of the
previously known short-period CVs: their SDSS spectra are dominated by
emission from the white dwarf, no spectroscopic signature from the
companion star is evident at optical wavelengths, and very few exhibit
obvious accretion activity such as X-ray emission or dwarf nova
outbursts. All these characteristics suggest that these systems have
very low accretion rates, and they are most likely all WZ\,Sge-type
dwarf novae with extremely long recurrence times. Among the
  white-dwarf dominated CVs discovered by SDSS, SDSS\,J0804+5103 is so
  far the only one to have revealed itself as a WZ\,Sge star
  \citep{pavlenkoetal07-1}.

\section{Discussion}
\label{s-discussion}

We have shown in Sect.\,\ref{s-period_dist} that the period
distribution of the SDSS CVs differs dramatically from that of the
previously known CVs, with a substantially larger fraction of
below-the-gap to above-the-gap systems, and a significant accumulation
of CVs at the orbital period minimum. More specifically, the origin of
the 80--86\,min period spike is entirely due to the new CVs identified
in SDSS, and, as outlined in Sect.\,\ref{s-spike_prop}, the systems in the
period spike differ also in spectral morphology and accretion activity
from the longer period CVs.  Here, we will discuss why the CVs in the
SDSS sample differ so clearly from the previously known systems, in
particular the effects of survey depth and CV candidate selection.

\subsection{Deep, deeper, the deepest}
\label{s-deep}
One very obvious difference between the SDSS CVs and CV samples from
previous surveys is the unrivalled depth of Sloan. Hence, SDSS should
be able to identify systems that were intrinsically too faint, or at
too large a distance, for the previous surveys.  This raises the
question \textit{does SDSS find more short period CVs than previous
  surveys, such as the Palomar Green (PG) Survey or the Hamburg Quasar
  Survey, just because of its depth?} A full treatment of this
question would require the computation of a galactic model of the CV
population, which would need to be folded through the details of the
sky coverage, magnitude limit, and colour cuts of the considered
surveys (see \citealt{pretoriusetal07-1} for an analysis of this kind
for the Palomar Green survey). Given the intricate allocation
algorithms for spectroscopic fibres within SDSS, this task is beyond
the scope of the present paper (see, however, Sect.\,\ref{s-exbox} for
a brief discussion on the colour-colour exclusion boxes of the quasar
target algorithm).  For simplicity, we develop here an empirical
comparison between SDSS, the Hamburg Quasar Survey (HQS), and the PG
survey, using their \textit{effective survey volumes} for the
white-dwarf dominated systems near the period minimum.  The effective
survey volume is calculated by integrating over a spherical cap in
galactic coordinates covering galactic latitudes higher than
$|b_\mathrm{lim}|$, weighting the volume by an exponential drop-off in
the space density of CVs along the $z$-axis, with a scale height
$H_z$. We assume radial symmetry around the $z$-axis in both the
distribution of CVs and the coverage of the survey. Any more realistic
work would need to account for the dependence of the CV space density
on galactic longitude, plus the exact tiling of the different surveys,
as none of them covers the full spherical cap. Finally, the effective
volume is weighted by the survey area.

The scale height is a poorly determined parameter, and we will assume
in the following two values for $H_z$. The first one is the canonical
value $H_z=190$\,pc introduced by \citealt{patterson84-1}, broadly
supported by \citet{thomas+beuermann98-1}, who found $H_z=155$\,pc for
strongly magnetic CVs, and \citet{aketal08-1} who found
$H_z=128-160$\,pc. The second value adopted below is $H_z=260$\,pc,
following the argument of \citealt{pretoriusetal07-1} that old
(short-period) CV populations are expected to have a larger scale
height than younger objects. \citet{vanparadijsetal96-1} determined
$H_z=160-230$\,pc from an analysis of the systemic velocity
distribution of CVs, and concluded that CVs are an old disk
population, with a mix of ages up to
10\,Gyr. \citet{pretoriusetal07-1} adopt $H_z=450$\,pc for period
bouncers, being the oldest CVs, however, given the fact that there is
little evidence of a period-bounce population in the SDSS CV sample,
we will not make use of such a large value for $H_z$.

We start with an estimate of the absolute magnitudes of the period
minimum CVs, which we need to turn the magnitude limits of the surveys
into distance limits, followed by a brief summary of the survey
characteristics, and then delve into the actual comparison of their
results in terms of CV discoveries.

\begin{table*}
\caption{Comparison of the SDSS, HQS, and Palomar Green Survey in
  terms of their potential for discovering period-minimum CVs.  The
  first three columns give the magnitude limit, sky coverage, and
  galactic latitude range for the three surveys. For SDSS, we defined
  subsamples with different limiting magnitudes.
  $g_\mathrm{lim}=22.5$ corresponds to the full CV sample,
  $g_\mathrm{lim}=19.0$ includes all CVs within the magnitude limit of
  the low-redshift quasar survey, and $g_\mathrm{lim}=17.4$ and
  $g_\mathrm{lim}=16.1$ are selected to compare SDSS like-for-like
  with the HQS and the PG survey.  From these characteristics, we
  calculated an effective survey volume for white-dwarf dominated CVs
  as outlined in Sect.\,\ref{s-deep}. The survey volumes were
  calculated for two different assumptions of the scale height, $H_z$,
  of the CV population, and normalised to the volume of the SDSS
  low-redshift quasar survey ($g_\mathrm{lim}=19.0$). The last two
  columns give the number of all CVs with $\Porb\le86$\,min found in
  the three surveys, and the number of period-minimum CVs with
  white-dwarf dominated spectra. Numbers following a "/" are
  normalised to the SDSS values.}
\label{t-survey_comparison}
\begin{tabular}{cccccccccc}\hline
Survey &
$g_\mathrm{lim}$ &
$d_\mathrm{lim}$ &
Area & 
$|b|$ & 
\multicolumn{2}{c}{$V_\mathrm{norm}$} &
\multicolumn{2}{c}{$N_\mathrm{CV}(\Porb\le86\,\mathrm{min})$} & \\ 
&& [pc] & $[\deg^2]$ && $H_z=190$\,pc &  $H_z=260$\,pc &
\multicolumn{1}{c}{all CVs} & \multicolumn{1}{c}{wd-dominated} \\\hline
SDSS & 22.5 & 1514 & 6400  & $>30^\circ$ & 1.36 & 4.11 & 33/1.57 & 20/1.81\\
     & 19.0 &  302 &       &             & 1.00 & 1.00 & 21/1.00 & 11/1.00\\
     & 17.4 &  145 &       &             & 0.24 & 0.20 &  5/0.20 & 0/0.00 \\
     & 16.1 &   79 &       &             & 0.05 & 0.04 &  1/0.0  & 0/0.00 \\\hline
HQS  & 17.4 & 145 & 13600 & $>20^\circ$ & 0.50 & 0.40 &  6/0.27 &  1/0.08 \\
PG   & 16.1 &  79 & 10714 & $>30^\circ$ & 0.08 & 0.05 &  3/0.14 & 1:/0.08 \\
\hline
\end{tabular}
\end{table*}

\subsubsection{Absolute magnitudes of the period minimum cataclysmic variables}
\label{s-absmag}
Unfortunately, the absolute magnitudes of CVs are notoriously poorly
determined, as there are too few systems with accurate distance
determinations to carry out a reliable calibration. With this caveat
in mind, we will now compare the absolute magnitudes of the
white-dwarf dominated CVs found by SDSS to those of the previously
known systems with $\Porb\le86$\,min.

Among the 7 old SDSS CVs with $\Porb\le86$\,min
(Table\,\ref{t-prevknown}), there is just one system with a
trigonometric parallax, the dwarf nova T\,Leo
\citep{thorstensen03-1}. From $d=101{+13\atop-11}$\,pc and $g=14.9$ we
find $M_g=9.9\pm0.3$. For SW\,UMa and EG\,Cnc spectral modelling of
the white dwarf in ultraviolet \textit{HST}/STIS resulted in distance
estimates of $d=159\pm22$\,pc and $d=420\pm65$\,pc
\citep{szkodyetal02-3, gaensickeetal05-2}, respectively, which results
in $M_g=10.9$ and $M_g=10.8$, respectively. Taking the average over
these three systems, we find $<M_g>=10.5\pm0.5$\footnote{For
  completeness, the polars EV\,UMa and GG\,Leo have lower limits on
  their distances, $d\ga705$\,pc and $d>100$\,pc
  \citep{osborneetal94-1, burwitzetal98-1}, which give $M_g\la7.3$ and
  $M_g\la12.2$, respectively. However, such lower limits on distances
  are very uncertain due to the possible contamination by cyclotron
  emission, and we do not include these two systems in our analysis
  above.}.

Distance estimates are available for 13 of the white-dwarf dominated new
SDSS CVs in the period-minimum spike: six from modelling high-speed
light curves \citep{littlefairetal08-1}, four from modelling the
optical SDSS spectra \citep{gaensickeetal06-1, mukadametal07-1}, and
three from modelling combined optical SDSS plus ultraviolet
\textit{HST}/ACS spectra \citep{szkodyetal07-1}. The resulting
absolute magnitudes range from $M_g=10.5-13.1$, with an average of
$<M_g>=11.6\pm0.7$. The large spread is likely to be caused by the
systematic uncertainties inherent to the distance determinations,
rather than substantial intrinsic differences in the system
properties. An independent check on these values can be obtained by
considering that the white dwarfs in these systems have typically
$\Teff\simeq12000-15000$\,K \citep{gaensickeetal06-1, mukadametal07-1,
  littlefairetal08-1}. Assuming an average mass of
$0.85$\,\Msun\ \citep{littlefairetal08-1}, and using an updated
version of the photometric white dwarf calibrations by
\citet{bergeronetal95-2}, we obtain $M_g=12.2-11.8$ for the white
dwarfs alone. Given that, by definition, the white dwarf is dominating
the optical spectrum in these systems, adding a maximum of 50\%
accretion luminosity to the brightness results in $M_g=11.3-11.7$ for
the bulk of the new SDSS CVs in the period minimum spike, entirely
consistent with our estimate of $<M_g>=11.6\pm0.7$.

While not statistically significant, the absolute magnitudes derived
above suggest that the new SDSS CVs are on average intrinsically
fainter than the old SDSS CVs, which is not too surprising as the mere
fact that the white dwarf is the dominant source of light implies that
the accretion luminosity is low. It is interesting to compare our
numbers here with the work of \citet{patterson98-1}, who estimated
time-averaged absolute magnitudes for a large number of CVs, finding a
nearly flat distribution between $M_V=4-11$, with a sharp cut-off for
fainter systems. The faintest bin in Patterson's (1998) Fig.\,2 is
populated only by a handful of WZ\,Sge type dwarf novae, and it is in
that bin that the white dwarf dominated CVs (which are presumably all
WZ\,Sge type dwarf novae) will slot in. However, Patterson's
(\citeyear{patterson98-1}) statement \textit{"...with not a single
  star fainter than $<M_v>=11.6$"} still holds, as no system
significantly fainter than that limit has been found by SDSS.

\subsubsection{Sloan Digital Sky Survey} 
The SDSS covers high galactic latitudes, $|b|>30^\circ$
\citep{yorketal00-1}. Within the main quasar search, spectroscopic
follow-up is carried out on point-sources with colours different from
those of main sequence stars and a limiting magnitude of a (galactic
column) de-reddened $i=19.1$ for ultraviolet excess / low-redshift
quasars and of a de-reddened $i=20.2$ for high-redshift ($z\ga3$)
quasars \citep{richardsetal02-1}. The detailed fibre allocation
algorithm is complex, and we will assume for the moment that the
completeness in the follow-up of blueish CVs within the magnitude
limits is as high as that for the targeted ultraviolet excess quasars
($\sim90$\%, \citealt{schneideretal07-1}, see Sect.\,\ref{s-exbox}
below). Further, we assume a typical galactic reddening of
$E(B-V)=0.05$, which translates into a reddening correction in $i$ of
0.1\,mag, increasing the observed magnitude limit of the low-redshift
quasar survey to $i=19.2$. The white-dwarf dominated CVs have on
average $<g-i>\simeq-0.2$, which leads to a limiting magnitude for
such systems within the main quasar survey of $g\simeq19.0$. Using
$<M_g>=11.6$ as determined above, this implies that SDSS should be
able to serendipitously identify white-dwarf dominated systems out to
$d=302$\,pc. Given that \citet{szkodyetal07-2} published some CVs
beyond DR5, and we include here in addition to her lists
(\citealt{szkodyetal02-2} to \citealt{szkodyetal07-2}) the previously
known CVs within DR6, we assume a survey area for the spectroscopic
SDSS data base of $6400\,\deg^2$, which is in between the official DR5
and DR6 areas.

\subsubsection{Hamburg Quasar Survey}
The Hamburg Quasar Survey (HQS) is another high-galactic latitude
($|b|>20^\circ$) survey covering $13\,600\,\deg^2$ with a typical
limiting magnitude of $B=17.5$.  Using the colour transformations of
\citet{jesteretal05-1}, the HQS has a limiting magnitude of
$g\simeq17.4$ for blue objects. Spectroscopic follow-up over the
wavelength range 3400--5400\,\AA\ was obtained by means of Schmidt
prism spectroscopy, which is hence complete except for plate artifacts
or blends. About 50000 blue objects with $U-B\la-0.5$ were extracted
from the photographic plates and visually classified.  Objects with
Balmer emission lines were selected as CV candidates for detailed
follow up \citep{gaensickeetal02-2, aungwerojwitetal06-1}.  Given the
average colour of the white-dwarf dominated CVs found in SDSS of
$<u-g>\simeq0.15$, the HQS should be able to identify such systems
out to $d=145$\,pc.

\subsubsection{Palomar Green Survey}
The Palomar Green Survey (PG) extended over $10\,714\,\deg^2$ at
galactic latitudes $|b|>30^\circ$ with spectroscopic follow-up for
1874 objects \citep{greenetal86-1}. The survey design was a blue-cut
of $U-B<-0.46$ and a typical limiting magnitude of
$B<16.1$. Comparison with SDSS data in overlapping areas showed
however that the PG survey had a rather bluer cut of $U-B<0.71$
\citep{jesteretal05-1}. Using the colour transformation from
\citet{jesteretal05-1}, the PG colour cut and limiting magnitude are
$u-g<0.3$ and $g=16.1$. Hence the PG survey should be able to identify
short-period CVs out to $d=78$\,pc.

\subsubsection{Finding period-minimum cataclysmic variables: A comparison of SDSS, HQS, and PG}
Using the survey characteristics summarised above, we calculated the
effective survey volumes for finding period-minimum CVs, as outlined
above, for the SDSS, HQS, and PG survey. As we are only interested in
the relative ``catchment area'', we normalise all numbers to the
effective volume of the main quasar survey within SDSS, i.e. to SDSS
with a limiting magnitude of $g=19.0$, and report the resulting
numbers in Table\,\ref{t-survey_comparison}. Along with the normalised
effective volumes, we list in Table\,\ref{t-survey_comparison} the
number of CVs with $\Porb\le86$\,min identified by each survey, and in
a separate column the number of those period-minimum CVs in which the
white dwarf is detected in their optical spectra. For the case of
SDSS, we give three additional subsamples with the following limiting
magnitudes: (a) $g_\mathrm{lim}=22.5$, which is the magnitude of the
faintest SDSS CV, resulting in the most inclusive sample of systems,
and (b) $g_\mathrm{lim}=17.4$ and $g_\mathrm{lim}=16.1$, which will
allow investigation of how SDSS compares with the HQS and the PG
survey if operating at the same limiting magnitude.

The first thing to note is that the difference in effective survey
volumes depends only mildly on the different assumption for the scale
height, the reason being that both HQS and PG are so shallow that they
do not even extend beyond one scale height, and hence do not feel much
of the exponential drop-off in the CV space density. 

SDSS beats the HQS in terms of survey volume only by about a factor
two, which is due to the HQS covering more than twice the area on the
sky, and extending down to lower galactic latitudes. While the number
of period-minimum CVs found in the HQS is only slightly below the
expectation from simply scaling the survey volume, it has only
produced one white-dwarf dominated CV (V445\,And,
\citealt{araujo-betancoretal05-1}), which is far below the expectations
from its survey volume, suggesting a selection effect against the
detection of such systems. \citet{gaensickeetal02-2} showed that the
HQS is very efficient in finding short period CVs similar to those
known in the late 1990s, if they had H$\beta$ equivalent widths in
excess of $\sim10$\,\AA. Those were all CVs with substantial
accretion luminosity such as SW\,UMa or T\,Leo, as only very few
white-dwarf dominated CVs were known at that time. However,
\citet{gaensickeetal02-2} noted that the results from the HQS
\textit{``...exclude the presence of a large population of nearby
  infrequently outbursting X-ray faint short-period CVs unless they
  have significantly weaker emission lines than, e.g., WZ\,Sge''}. It
turns out that those types of systems, short-period CVs with no or
very rare outbursts, no or very weak X-ray emission, and weak H$\beta$
equivalent widths are frequent among the CV catch of SDSS
(Fig.\,\ref{f-compspec}). The difficulty in finding that type of
system in the HQS was exacerbated by the low spectral resolution,
averaging over the broad white dwarf absorption lines and the weak
emission lines, and thus even further decreasing the net equivalent
widths of the Balmer emission.

Comparing SDSS and the PG survey, the numbers of short-period CVs found
are roughly in line with the expectations from the scaled survey
volumes~--~only three CVs with $\Porb\le86$\,min were found, of which
one may be white-dwarf dominated.

A final point of our comparison is to inspect whether, with regard to
finding CVs near the period-minimum, SDSS is a superset of the three
surveys under inspection. The six CVs with $\Porb\le86$\,min in the
HQS are SW\,UMa, T\,Leo, DW\,Cnc and HT\,Cam, all of which were
previously known CVs with substantial accretion activity
(outbursts/X-ray emission), and the two new discoveries KV\,Dra
(HS1449+6415), an SU\,UMa type dwarf nova with rare outbursts and weak
X-ray emission \citep{jiangetal00-1, nogamietal00-1}, and V455\,And
(HS2331+3905), the only white-dwarf dominated period-minimum CV in the
HQS \citep{araujo-betancoretal05-1}. The first superoutburst of
V455\,And was observed in September 2007, confirming it as a WZ\,Sge
type dwarf nova with a superoutburst cycle $>5$ years. Of those six
period-minimum systems in the HQS only T\,Leo, SW\,UMa, DW\,Cnc are in
the footprint of SDSS DR6, and all were spectroscopically followed-up
by SDSS and, hence, (re-)identified as CVs.

The PG survey contains three systems with $\Porb\le86$\,min: the
previously known T\,Leo, and the PG discoveries RZ\,LMi and MM\,Hya
\citep{greenetal82-1}. RZ\,LMi is an SU\,UMa star with an ultra-short
outburst cycle \citep{robertsonetal95-1, nogamietal95-1}, which is
thought to reflect an unusually high mass transfer rate for its orbital
period \citep{osaki95-1, osaki95-2}.  MM\,Hya is a relatively poorly
studied dwarf nova with rather rare outbursts \citep{ringwald93-2,
misselt+shafter95-1}. The spectrum published by
\citet{zwitter+munari96-1} is of low quality, but suggests that
MM\,Hya may be a white-dwarf dominated CV similar to those discovered
in large number by SDSS. T\,Leo and RZ\,LMi are in the footprint of
SDSS DR6, but only T\,Leo has been followed up
spectroscopically. RZ\,LMi was found in the SDSS imaging at $g=14.6$,
close to the bright end where SDSS does follow-up observations, and
was rejected by the quasar target selection algorithm. 

In summary, comparing the numbers of period-minimum CVs found in SDSS,
the HQS, and the PG survey with the normalised effective survey
volumes shows that the three surveys produce broadly consistent
results. The main gain that SDSS brings over the previous surveys
comes from its depth, and the massive spectroscopic follow-up of
CV (and quasar) candidates.

\subsection{Caveats?}

Here, we discuss a number of possible caveats that could affect our
conclusions about the period distribution of SDSS CVs made above.

\begin{figure*}
 \includegraphics[angle=-90,width=18cm]{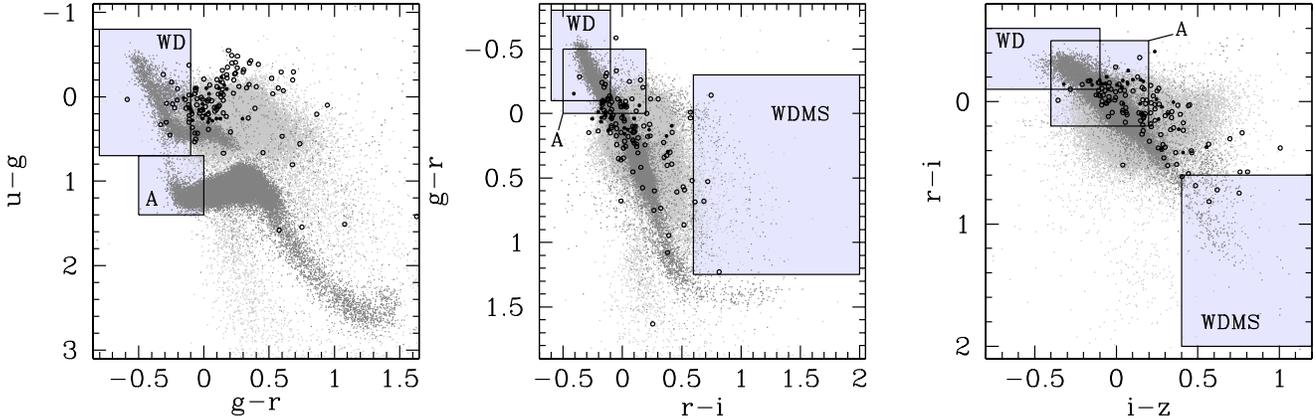}
 \caption{\label{f-colours} White dwarf (WD), A-star, and white dwarf
   plus main sequence (WDMS) binary $ugriz$ exclusion boxes in the
   spectroscopic follow-up of SDSS quasar candidates. To be excluded
   from spectroscopic follow-up, an object must be located in all
   three (WD, A-star) or two (WDMS) boxes. The distribution of single
   white dwarfs and main sequence stars is shown by dark gray dots,
   that of quasars by light gray dots. The SDSS CVs from
   Tables\,\ref{t-porbs} and~\ref{t-prevknown} are shown in black,
   with the filled circles representing the white-dwarf dominated CVs.}
\end{figure*}

\subsubsection{The exclusion boxes in the SDSS quasar target selection}
\label{s-exbox}

The decision on whether or not a photometric SDSS object will be
allocated a fibre for spectroscopic follow-up is very complex, as a
variety of science programs share the available resources. Because of
their composite nature (white dwarf, companion star, accretion flow),
the colours of CVs in $ugriz$ space differ markedly from those of
single main sequence stars or white dwarfs. A detailed analysis of the
spectroscopic completeness of SDSS as a function of location in
$ugriz$ colour space and apparent magnitude is in preparation and will
be presented elsewhere. Here, we focus on the question \textit{could
  the exclusion boxes in the SDSS quasar target selection algorithm
  cause a significant bias in the composition of the SDSS CV sample?}
Early during the operations of SDSS, it became clear that the
population of ultraviolet-excess quasars overlaps in $ugriz$ colours
space with white dwarfs, resulting in a substantial stellar
contamination of the quasar sample in these regions
\citep{richardsetal02-1}. Three $ugriz$ exclusion boxes were defined
to suppress the contamination of white dwarfs, A-stars, and white
dwarf plus main sequence (WDMS) binaries. Figure\,\ref{f-colours}
shows these colour exclusion boxes along with locations of single
stars, quasars, and of the CVs from Tables\,\ref{t-porbs} and
\ref{t-prevknown}. While excluded from the quasar candidate follow-up,
a sufficient number of objects within these colour boxes were still
observed within other SDSS science programs to establish their typical
nature. The only SDSS CVs discussed in this paper contained in the
white dwarf exclusion box are SDSS2116+1134 and SDSS0310--0755
  (none of which is white-dwarf dominated), which is a simple
consequence of the fact that Balmer emission lines move CVs away from
the colour locus of white dwarfs, even if their continuum flux is
dominated by the white dwarf. No CV is located in the A-star exclusion
box, and the WDMS binary exclusion box contains SDSS0751+1444
(PQ\,Gem), SDSS0808+3131, SDSS0900+4301, SDSS0938+0714 (PG\,0935+075),
and SDSS1554+2721. All five systems have $\Porb>150$\,min, and are
hence not included in the discussion in Sect.~\ref{s-period_dist}
and~\ref{s-spike_prop}. Their location in the WDMS exclusion box is
explained by the strong contribution of their companion stars (or, in
the case of PQ Gem, a field M-dwarf included in the SDSS fibre). It
may hence be that long-period CVs with low accretion rates
(i.e. visible companion star) may be under-represented in the SDSS CV
sample.  We conclude that the exclusion boxes in the SDSS quasar
follow-up have no noticeable effect on the conclusions drawn here.

\subsubsection{Bias in the follow-up strategy}

An obvious question is \textit{have we biased our follow-up in a way
that would favour observations of short-period CVs}? If such a bias
existed, the orbital period distribution resulting from the follow-up
work would be skewed even within the SDSS CV sample, independent of
the question of how representative this sample is of the true galactic
CV population. As shown in Sect.\,\ref{s-spike_prop}, a large fraction
of the systems in the period spike have white-dwarf dominated optical
spectra. Among the remaining $\sim130$ SDSS CVs with no accurate period
determination, the fraction of white-dwarf dominated systems is
similar to that within the sample of well-studied systems discussed
here, suggesting that the period distribution of these remaining
systems should be overall similar to the one shown here. In addition,
at least the earlier follow-up studies of SDSS CVs focused on
brighter systems, for the obvious reason of them being easier
observational targets. Hence, the early follow-up was more likely to
target intrinsically bright systems, which implies relatively high
accretion rates, and on average periods above the spike. Finally, as
it is evident from Fig.\,\ref{f-porbdist} (right panel), the 45
previously known CVs with accurate periods \textit{do not} show a
period minimum spike. Among the $\sim130$ SDSS CVs with no accurate
period, there is only a handful of additional previously known
systems, and one might expect that the fraction of systems within the
period spike will be higher in the remaining new SDSS discoveries. In
summary, we are confident that our follow-up strategy has not
introduced a significant bias with respect to the orbital periods of
the systems studied so far. However, given that the fraction of
well-studied systems among the SDSS CV sample is still relatively
small ($<50$\%), further follow-up work is highly desirable to improve
the statistics of the results presented here.

\begin{figure*}
 \includegraphics[angle=-90,width=\columnwidth]{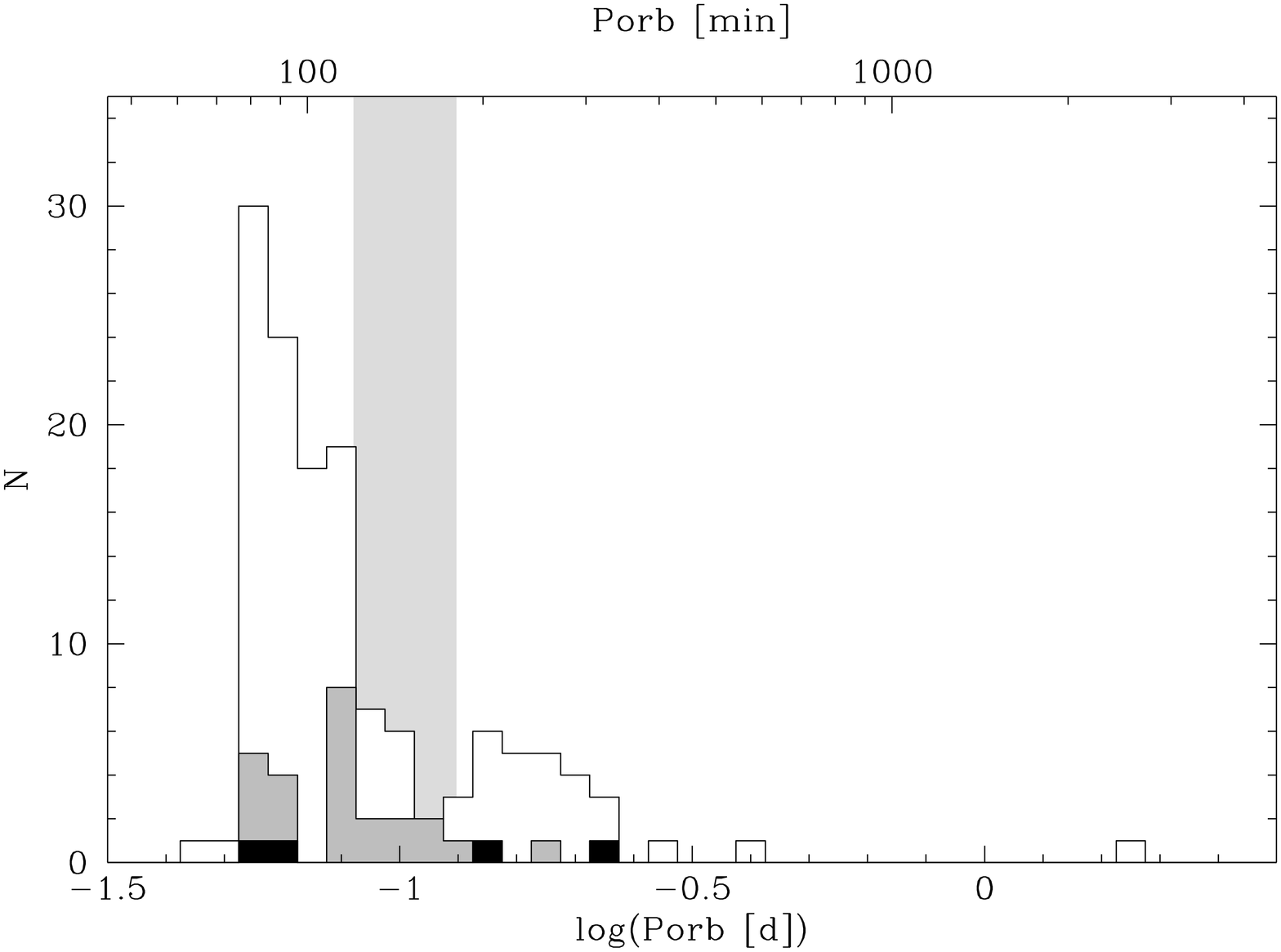}
 \includegraphics[angle=-90,width=\columnwidth]{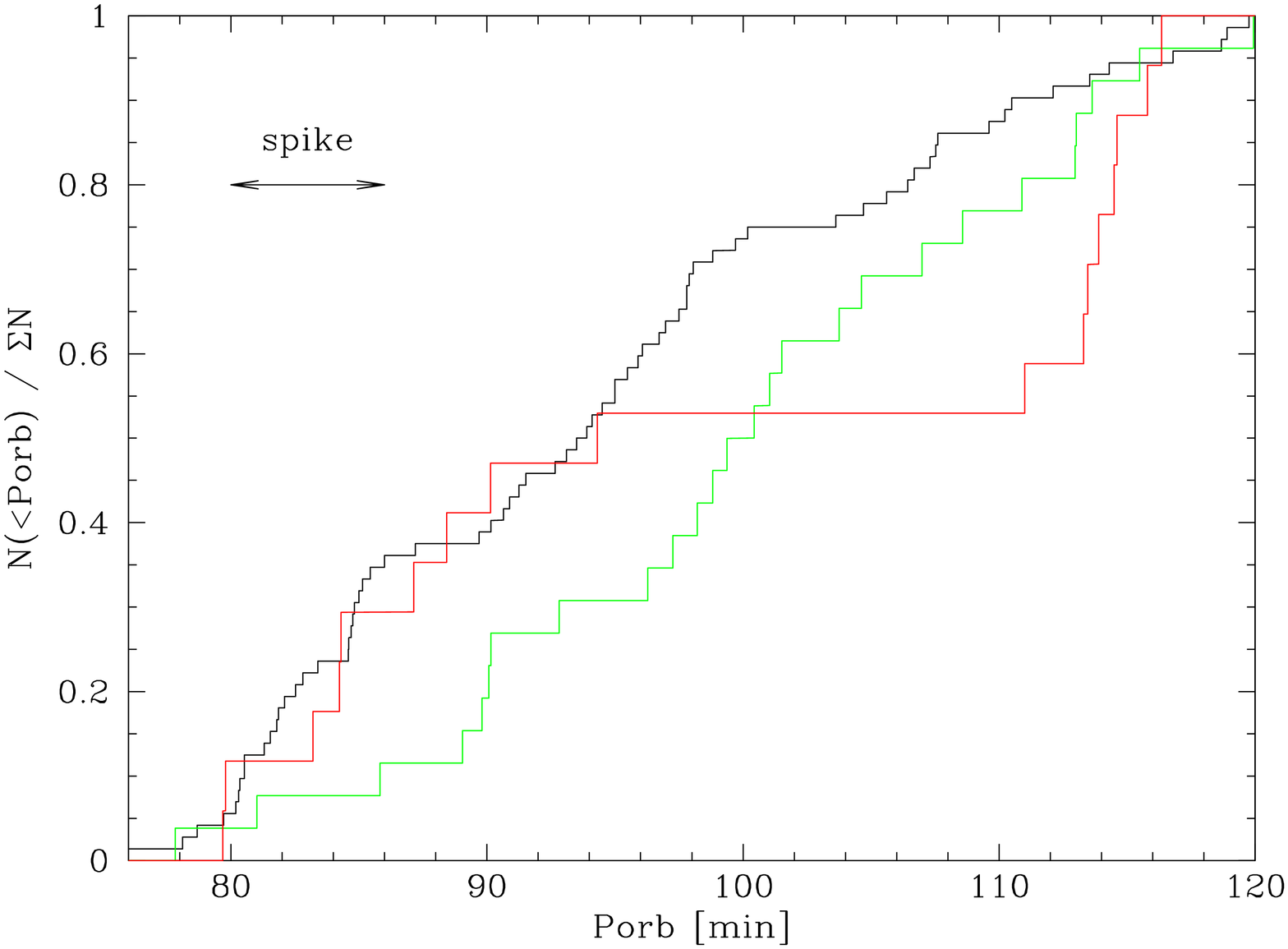}
 \caption{\label{f-mcv_vs_nmcv} Left panel: The orbital period
   distributions of non-magnetic SDSS CVs (Table\,\ref{t-porbs},
   \ref{t-prevknown}, white histograms), the 25 polars among the SDSS
   CVs (gray) and the 4 IPs among the SDSS CVs (black). Right panel:
   cumulative period distributions of the SDSS CVs (short dashed), the
   17 polars among the SDSS CVs with $\Porb\le120$\,min (dotted) and
   54 polars among the non-SDSS CVs (solid). The distribution of the
   polars among the SDSS CVs features an accumulation of systems
   around 114\,min, which is predominantly made up of previously known
   systems. This ``spike'' has been suggested to be the signature of
   the resumption of mass transfer at the lower edge of the period gap
   \citep{hameuryetal90-1}, but this interpretation is doubtful in
   light of the much more detailed period distribution of polars
   available today.}
\end{figure*}

\subsubsection{Magnetic CVs}

Another caveat to bear in mind is that the intrinsic and observed CV
populations may be made up of different evolutionary channels. It has
been suggested that the strong magnetic field in polars will reduce
the rate at which magnetic braking extracts angular orbital momentum
\citep{webbink+wickramasinghe02-1}, which would result in different
evolutionary timescales (and hence different period distributions)
between non-magnetic and strongly magnetic CVs. In the standard
scenario of CV evolution, magnetic braking ceases at
$\Porb\simeq3$\,h, and hence the populations of polars and
non-magnetic CVs should differ most dramatically above the period
gap. Plus, as polars do not suffer a discontinuous change in angular
momentum loss, they are not expected to exhibit a period gap.  One
piece of observational evidence supporting the hypothesis that polars
above the period gap have lower mass transfer rates, and hence are
likely subject to lower rates of angular momentum loss, is that the
white dwarfs in polars have consistently lower effective temperatures
than those in non-magnetic CVs \citep{sion91-1, townsley+bildsten03-1,
  araujo-betancoretal05-2}. A similar behaviour is observed below the
period gap, though at a much lower level
\citep{araujo-betancoretal05-2}, which may suggest that non-magnetic
CVs below the period gap are subject to some amount of residual
magnetic braking, causing slightly higher mass transfer rates than
found in short-period polars. Such differences in angular momentum
loss rates should result in different orbital minimum periods for
polars and non-magnetic CVs~--~which is, however, not observed. For
completeness, we show in Fig.\,\ref{f-mcv_vs_nmcv} the binned and
cumulative period distributions of polars in the SDSS CVs and the
non-SDSS CV samples. The fraction of polars in the SDSS CV sample is
$\sim20$\%, compatible with the fraction among all known CVs
\citep{wickramasinghe+ferrario00-1}.

Formally, a KS test on the cumulative period distributions of the
non-magnetic SDSS CVs and the SDSS polars results in a 3.2\%
probability for both sets being drawn from the same parent sample,
which seems to suggest an intrinsic difference between these two
samples. Comparing the polars among the non-SDSS CVs and the SDSS
polars, the probability that both sets are drawn from the same parent
population is 27\%, i.e. the two samples are broadly consistent in
their period distribution.  Taking these numbers at face value, the
SDSS polars appear to have a different period distribution from the
non-magnetic SDSS CVs.  Eye-balling the distributions in
Fig.\,\ref{f-mcv_vs_nmcv}, the accumulation of SDSS polars around
114\,min is striking, and is in fact the dominant culprit for the
statistical differences between the three CV samples inspected
here. Tables\,~\ref{t-porbs} and \ref{t-prevknown} reveal that this
accumulation is predominantly made up of the previously known systems,
which were all identified as CVs because of their X-ray emission. The
significance of this ``114\,min spike'' has been debated when a total
of only 15 polars were known, and has been suggested to be related to
the resumption of mass transfer at the lower end of the period gap
\citep{hameuryetal90-1}. In the period distribution of all $>80$ polars,
selected in the vast majority of cases from X-ray surveys, this spike
is gone. We conclude that this feature, when identified 20 years ago,
served as a powerful demonstration of small number statistics, and
fulfils the same job a second time around, as it happens that most of
the long-known ``114\,min'' polars are in the SDSS footprint, whereas
many of the polars found since then, e.g. in the ROSAT All Sky Survey,
are not.

\subsubsection{Thermal timescale mass transfer CVs}

Another channel that is likely to add noticeably to the CV population
are systems which started with a mass ratio
$M_\mathrm{donor}/M_\mathrm{wd}\ga1$, and underwent a phase of thermal
time-scale mass transfer (TTMT) before evolving into CVs
\citep{schenkeretal02-1, podsiadlowskietal03-1, kolb+willems05-1}. A
number of suspected post-TTMT systems have been found
\citep[e.g.][]{thorstensenetal02-2, thorstensenetal02-1,
  gaensickeetal03-1}, but again their number is too small to assess
their effect on the overall orbital period distribution of CVs.

\subsection{Implications on cataclysmic variable population models}

From our follow-up studies of CVs discovered by SDSS we have
identified the ``spike'' at the minimum period predicted by theory for
two decades. Bearing in mind the caveats outlined above, what other
implications can we derive from the SDSS CV sample at the current
stage of follow-up, with only about half of the systems having
reliable orbital periods?  

\subsubsection{Angular momentum loss rates below the period gap}
Standard population models with gravitational wave radiation as the
only driver of mass transfer in CVs below the period gap place the
period minimum spike near 70\,min \citep{kolb+baraffe99-1,
  barker+kolb03-1}. Tidal and rotational corrections of the underlying
one-dimensional stellar models do not affect the position of the
systems' period bounce, a finding supported by a 3-dimensional SPH
model of Roche-lobe filling stars \citep{renvoizeetal02-1}.  Our
discovery of a period spike near 80\,min strongly favours the view
that the observed CV period minimum is indeed the result of a period
bounce at 80\,min, and that the theoretically calculated period
minimum is too short by about 10\,min.

The observed and calculated period minimum can be reconciled if either
the orbital braking is about four times the value provided by
gravitational wave radiation \citep{kolb+baraffe99-1}, or if the
theoretical models underestimate the stellar radius for a given mass
by about 20\% \citep{barker+kolb03-1}.  An immediate consequence of
the former interpretation is that the fraction of systems below the
period gap, the fraction of post-period bounce systems, and the space
density of CVs should be smaller than in the canonical model by e.g
\citep{kolb93-1}.

The absence of a period minimum spike plagued theorists working on
compact binary evolution for more than two
decades. \citet{kingetal02-1} showed that it is possible to obtain a
more or less flat period distribution from subpopulations with
different AM loss rates, up to five times the rate of gravitational
wave radiation and fine-tuned contributions to the full
population. Here, we have shown that a period minimum spike
exists. Fig.\,\ref{f-cumulative} suggests that the spike is
  limited by two breaks in the slope of the cumulative period
  distribution, located at 80\,min and 86\,min, defining an
  approximate width of $\simeq6$\,min. The period spike can be fit
  with a binned Gaussian model. We used a Monte-Carlo simulation
  assuming a Gaussian distribution to estimate $1\,\sigma$ errors on
  the parameters. This exercise results in the period of the spike,
  $P_\mathrm{spike}=82.4\pm0.7$\,min and its width
  $\mathrm{FWHM}_\mathrm{spike}=5.7\pm1.7$\,min. We stress that these
  values are only valid under the assumption of a Gaussian
  distribution, but do illustrate that a more detailed assessment of
  the spike structure will require a larger set of periods accurate to
  better than 1\,min. The sharp minimum period, coupled with the
finite width of the spike, strongly suggests that for the bulk of CVs
below the period gap there is very little variation in the secular
mean value of the AM loss rate for systems with similar
parameters. The width of the observed spike appears slightly larger
than typically predicted by population models ($\sim1.5-3$\,min),
however given that we requested for all CVs in the sample discussed in
this paper $\sigma(\Porb)/\Porb\le3\%$, the intrinsic width of the
period spike is not properly resolved, underlining the need for
accurate period measurements.

\subsubsection{The scale height of cataclysmic variables}
A simple statement on the scale height of CVs can be made simply from
comparing the SDSS subsamples for $g_\mathrm{lim}=19.0$, i.e. the CVs
from the main quasar sample (which we have assumed to be
approximately complete due to the colour-overlap between CVs and
quasars), and the full sample with $g_\mathrm{lim}=22.5$ (where SDSS
makes no attempt to achieve any level of completeness, but merely
allocates spare fibres to faint quasar candidates). A simple scaling
of the survey volumes predicts an increase in the number of CVs by
$\sim1.4$ for an assumed $H_z=190$\,pc. This modest gain comes from
the fact that, with a limiting distance of 302\,pc, the main quasar
survey at $g_\mathrm{lim}=19.0$ already extends out to $\sim1.6$ scale
heights, and not many more CVs are found by going further into the
halo. However, for an assumed $H_z=260$\,pc, the increase in the
number of CVs found by going to $g_\mathrm{lim}$ is expected to be
$\sim4$. Going to $g_\mathrm{lim}=22.5$ boosts the number of CVs
actually found by SDSS by $\sim1.5$, which exceeds the prediction for
$H_z=190$\,pc. This is a very conservative lower limit, as (a) the
spectroscopic follow-up of faint quasar candidates is very incomplete
down to $g_\mathrm{lim}=22.5$ \citep{richardsetal04-1}, and (b) there
are 16 more published WD-dominated systems without orbital period
determinations, of which 14 have $g>19.0$. Given the properties of the
systems shown in Fig.\,\ref{f-venn}, there is a $\sim80$\% probability
that those 16 systems will have periods $\Porb\le86$\,min as well. We
conclude from this empirical study that the scale height of CVs is
very likely larger than the 190\,pc used by \citet{patterson84-1}, in
agreement with the arguments of \citet{pretoriusetal07-1} and the
study by \citet{vanparadijsetal96-1}. 

\begin{figure}
 \includegraphics[width=\columnwidth]{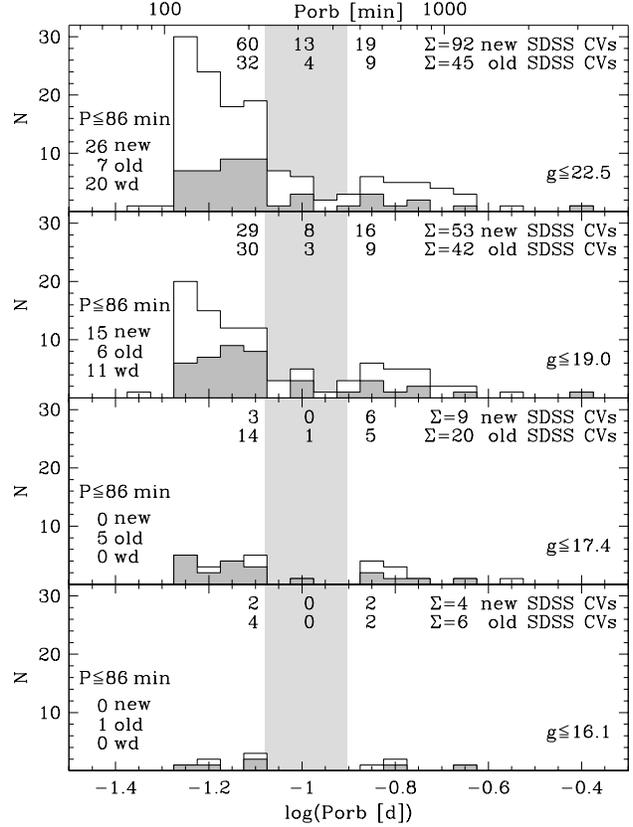}
 \caption{\label{f-magcut} The orbital period distributions of the new
   SDSS CVs (Table\,\ref{t-porbs}, white histograms) and the old SDSS
   CVs (Table\,\ref{t-prevknown}, gray) applying four different
   magnitude limits. From top to bottom: $g\le22.5$, corresponding to
   the faintest CV included in our sample; $g\le19.0$, corresponding
   to the magnitude limit of the low-redshift quasar survey within
   SDSS, $g\le17.5$ corresponding to the magnitude limit of the
   Hamburg Quasar Survey; and $g\le16.1$, corresponding to the
   magnitude limit of the Palomar Green Survey. The number of systems
   below, in, and above the 2--3\,h period gap as well as their sum
   are given in each panel, with the values for the new SDSS CVs being
   on top of those for the old SDSS CVs. The number of new SDSS CVs,
   old SDSS CVs, and white-dwarf dominated CVs in the $\sim80-86$\,min
   period spike are shown on the left of the period histogram.}
\end{figure}

\subsubsection{The space density of ``WZ\,Sge'' stars}

Assuming the standard CV evolution theory, and a mid-plane CV space
density of $5\times10^{-5}\mathrm{pc^{-3}}$, \citet{pretoriusetal07-1}
calculated the period distributions of various magnitude-limited
samples, and found that the period minimum spike decreases in
prominence for brighter magnitude limits, although it was still
present in a sample with $V<14$ with a contrast of two with respect to
neighbouring periods .

In Fig.\,\ref{f-magcut}, we show the period distributions of the SDSS
CVs, as before divided in new and old systems, for the four different
magnitude limits used in the previous sections, $g_\mathrm{lim}=22.5,
19.0, 17.4, 16.1$. The accumulation of systems at the minimum period
is clearly present in the full sample, and still so once a
$g_\mathrm{lim}=19.0$ cut is applied. At $g_\mathrm{lim}=17.4$, the
number of systems in the shortest-period bin still exceeds all other three
bins below the period gap, but the numbers are too small to draw any firm
conclusion. This being said, we note that a similar ``accumulation'' in
the shortest-period bin was found in the period distribution of HQS
CVs, which is an independent sample with the same magnitude limit (see
Fig.\,18 of \citealt{aungwerojwitetal06-1}). Cutting to
$g_\mathrm{lim}=16.1$, the limit of the PG survey, only 5 systems are
left below the period gap, and hence no statement with regard to a
period spike can be done.

\begin{figure}
 \includegraphics[angle=-90,width=\columnwidth]{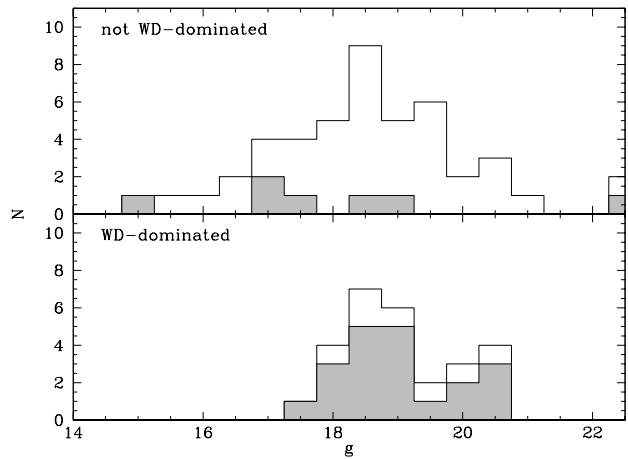}
 \caption{\label{f-wd_vs_nwd} Top panel: $g$-band magnitude
   distribution of non-magnetic SDSS CVs which do not exhibit broad
   photospheric white dwarf Balmer absorption lines in their optical
   spectra. Shown in white are all systems with $\Porb<120$\,min,
   i.e. below the period gap. Systems within the ``period minimum
   spike'', $\Porb\le86$\,min are shown in gray. Bottom panel: same as
   above, but for the white-dwarf dominated SDSS CVs (``WZ\,Sge
   candidates'').}
\end{figure}

From Fig.\,\ref{f-magcut}, it is clear that SDSS is picking up many
CVs with $\Porb\le120$\,min at magnitudes fainter than
$g_\mathrm{lim}\simeq17.4$. In particular, SDSS has found a large
number of CVs that have white-dwarf dominated optical spectra, which
must have very low-mass donor stars as little or no spectral features
from the companions are seen, and orbital periods right at the minimum
period. Taking their properties at face value, it appears likely that
all these systems are WZ\,Sge type dwarf novae, with very long
outburst recurrence times. There is no plausible selection mechanism
within the fibre allocation of SDSS that would go \textit{against}
finding this type of system at bright magnitudes $g\simeq15-17.4$, but
not a single bright WZ\,Sge candidate has been found by SDSS (the
brightest one being SDSS1339+4847, $g=17.6$).
Figure\,\ref{f-wd_vs_nwd} compares the magnitude distributions
of non-magnetic SDSS CVs with a white-dwarf dominated spectrum and
of those in which the white dwarf is not visible. Both distributions
peak at $g\simeq19$, i.e. the magnitude limit of the main quasar
sample, implying that the completeness of the SDSS CV sample drops
substantially at $g\ge19$. The white-dwarf dominated systems ramp up
in numbers only at faint magnitudes corresponding to distances
$d\ga150$\,pc, suggesting that they are intrinsically relatively
rare.

\begin{table}
\caption{\label{t-wzsge}WZ\,Sge candidate systems brighter than $V\simeq17$
  identified prior to SDSS}
\begin{center}
\begin{tabular}{lcccl}\hline
System    & $V_\mathrm{q}$ & $d\,[pc]$  & $M_V$ & Ref. \\\hline
WZ\,Sge   & 15.0 & $43\pm{1.6\atop1.5}$ & 11.8  & 1,2,3 \\
BW\,Scl   & 16.4 & $131\pm18$           & 10.8  & 4,5,6 \\
GD\,552   & 16.5 & $105\pm20$           & 11.4  & 7,8 \\
V455\,And & 16.5 & $74\pm{8\atop7}$     & 12.2  & 9,10 \\ 
GW\,Lib   & 16.7 & $104\pm{30\atop20}$  & 11.6  & 2,11 \\ \hline
\end{tabular}
\end{center}

$^1$\citet{mackie20-1},
$^2$\citet{thorstensen03-1},
$^3$\citet{harrisonetal04-3},
$^4$\citet{augusteijn+wisotzki97-1},
$^5$\citet{abbottetal97-1},
$^6$\citet{gaensickeetal05-2},
$^7$\citet{greenstein+giclas78-1},
$^8$\citet{undasanzanaetal08-1},
$^9$\citet{araujo-betancoretal05-1},
$^{10}$G\"ansicke et al. in prep,
$^{11}$\citet{gonzalez83-1} 
\end{table}

This suggestion is confirmed by looking at how many bright WZ\,Sge
(candidate) stars were known prior to SDSS. In Table\,\ref{t-wzsge},
we list the WZ\,Sge (candidate) systems from \citet{ritter+kolb03-1}
(V7.9) which are brighter than $V\simeq17$ in quiescence. The absolute
magnitudes for these five systems, $<M_V>=11.6\pm0.5$, agree well with
our estimate for the white-dwarf dominated SDSS systems
(Sect.\,\ref{s-absmag}). WZ\,Sge and GW\,Lib were discovered through
their large-amplitude outbursts, BW\,Scl, GD\,552, and V455\,And were
identified through follow-up spectroscopy of ROSAT X-ray sources, high
proper motion objects, and emission line stars from the HQS,
respectively. With the exception of GD\,552, all systems have
white-dwarf dominated spectra with no spectroscopic evidence for the
mass donors. Thus, it appears that no matter what selection method is
used, WZ\,Sge stars are not very numerous. 

The standard CV population models predict that $\sim99$\% of all CVs
should have periods $\la2$\,h, and $\sim70$\,\% should be post-bounce
systems \citep{kolb93-1, howelletal01-1}. If we follow
\citet{pretoriusetal07-1}, and take the average of the space densities
predicted by CV evolution theory,
$\rho=5\times10^{-5}\mathrm{pc^{-3}}$ \citep{dekool92-1,
  politano96-1}, we expect $\sim210$ CVs within a radius of 100\,pc of
the Earth, and practically all of them should be short-period systems,
of which $\sim150$ post-bounce. If the CV space density were as high
as theory predicts, comparing the numbers discussed above implies (a)
the bulk of CVs has still not been found, and (b) the majority of
period-bouncers must differ in terms of their colours and spectra from
the typical ``WZ\,Sge'' stars.

\section{Conclusions}
SDSS is providing us with the largest, deepest, and most homogeneously
selected CV sample so far, and holds the potential of detailed tests
of the current models of compact binary evolution. Follow-up studies
are currently available for about half of the SDSS CVs, and we have
analysed the global properties of this sample, with particular
emphasis on the numerous short-period systems. In summary, we find
that

\begin{enumerate}
\item The period distribution of the SDSS CVs differs significantly
  from that of any previous CV sample, containing a much larger
  fraction of short-period CVs. Most striking is the appearance of an
  accumulation of CVs with $80\la\Porb\la86$\,min, which we identify
  as the period-minimum spike predicted for two decades by CV
  population models. The finite width of the spike suggests that the
  angular momentum loss rates below the period gap must fall into a
  narrow range.
\item Within the period minimum spike, the majority of new CVs
  identified by SDSS differ substantially from the bulk of the
  previously known short-period CVs. Their optical spectra are
  dominated by the white dwarf, with no discernable signature from the
  mass donors, and only a few of them have so far exhibited
  ``typical'' CV accretion activity, i.e. X-ray emission or dwarf nova
  outbursts.  We determine an average absolute magnitude of
  $<M_g>=11.6\pm0.7$ for these systems. The spectra are usually devoid
  of spectroscopic signatures from the mass donors, implying very late
  spectral types and low masses. Work by \citet{littlefairetal06-2,
    littlefairetal07-1, littlefairetal08-1} confirms that a high
  fraction of the eclipsing systems in the period-minimum spike
  contain brown dwarf donors.
\item Comparison with the Hamburg Quasar Survey and the Palomar Green
  Survey suggests that the main advantage of SDSS is indeed its
  unprecedented depth. Comparing subsamples of the SDSS CVs with
  different magnitude limits indicates that the scale height of
  short-period CVs is $H_z>190$\,pc. 
\item While WZ\,Sge stars are often considered to be period-bounce
  candidate, the number of WZ\,Sge stars found so far is too small to
  account for the space density of post-bounce systems predicted by
  population models. 
\item Accurate period determinations should be obtained for the
  remaining $\sim130$ SDSS CVs, which will not only allow to put the
  results presented here on a more robust statistical base, but also
  to carry out more detailed investigations into the shape and width
  of the period minimum spike, and to address questions such as
  whether strongly magnetic CVs evolve differently from non-magnetic
  systems.
\end{enumerate}

\section*{Acknowledgements}
MD and JS acknowledge  STFC for a studentship and a PDRA grant. 
JT acknowledges NSF grants AST-0307413 and AST-0708810.
AA thanks the Royal Thai Government for generous funding.
PS acknowledges some support from NSF grant AST 02-05875.
SCCB was supported by FCT. 
MRS acknowledges support from FONDECYT (1061199).
The NASA Astrophysics Data System has been an important resource of
information for the research presented here. Funding for the SDSS has
been provided by the Alfred P. Sloan Foundation, the Participating
Institutions, the National Aeronautics and Space Administration, the
National Science Foundation, the US Department of Energy, the Japanese
Monbukagakusho and the Max Planck Society. The SDSS web site is
http://www.sdss.org. We would like to thank the amateur community
worldwide for their dedicated monitoring of the SDSS CVs. Finally, we
thank Joe Patterson for a constructive referee report.
       
\bibliographystyle{mn_new}                                              
\bibliography{aamnem99,aabib,proceedings,submitted}

\clearpage
\bsp
\label{lastpage}

\end{document}